\documentclass[12pt]{article}
\usepackage[round]{natbib}
\usepackage[hmargin=1in,vmargin=1in]{geometry}
\usepackage{latexsym,epsf,authblk}
\usepackage{url}
\usepackage{mathptmx}
\usepackage{amsfonts}
\usepackage{amssymb,amsmath,amsthm}
\usepackage[nooneline]{subfigure}
\usepackage{graphicx}
\usepackage{ifthen}
\usepackage{setspace}

\usepackage[printfigures]{figcaps}

\usepackage[none]{hyphenat}

\newboolean{publ}

\newcommand{\supp}{\textit{Supp. mat.}}

\theoremstyle{definition}

\theoremstyle{plain}

\newcommand{\ssm}{\mathtt{SSM}}

\newcommand{\kth}{\mathtt{K81}^{\ast}}
\newcommand{\ktw}{\mathtt{K80}^{\ast}}
\newcommand{\jc}{\mathtt{JC69}^{\ast}}

\newcommand{\ema}{\mathtt{EM}}
\newcommand{\empar}{\texttt{Empar}}
 
\renewcommand{\a}{\mathtt{A}}
\renewcommand{\c}{\mathtt{C}}
\newcommand{\g}{\mathtt{G}}
\renewcommand{\t}{\mathtt{T}}

\newcommand{\len}{\mathsf{L}_{\tree}}
\newcommand{\m}{\mathcal{M}}

\newcommand{\br}{\mathbf{l}}

\newcommand{\GMM}{\mathtt{GMM}}
\newcommand{\SSM}{\mathtt{SSM}}

\newcommand{\Kb}{\mathtt{K81}^{\ast}}
\newcommand{\JC}{\mathtt{JC69}^{\ast}}

\newcommand{\fish}{\textbf{I}}
\newcommand{\lik}{\mathcal{L}}
\newcommand{\tree}{\tau}

\newcommand{\edges}{E(\tree)}
\newcommand{\leaves}{{L(\tree)}}      
\newcommand{\me}{{A}^{e}} 

\newcommand{\estep}{\emph{E-step}}
\newcommand{\mstep}{\emph{M-step}}

\newcommand{\sone}{\tree_{\mathrm{balanced}}}
\newcommand{\stwo}{\tree_{1:2}}
\newcommand{\sth}{\tree_{2:1}}

\newcommand{\eltwo}{\mathsf{L}_2}  

\newcommand{\Keywords}[1]{\par\noindent {\small{\em Keywords\/}: #1}}

\title{$\empar$: EM-based algorithm for parameter estimation of Markov models on trees.}

\author{A. M. Kedzierska$^{1,2}$}
\author{ M. Casanellas$^{2,**}$ }
\affil{ 1. Bioinformatics and Genomics Group, Center for  Genomic Regulation\\
Barcelona Biomedical Research Park (PRBB)\\
c/ Dr. Aiguader, 88, 08003 Barcelona
}
\affil{2. Dpt. Matem\`{a}tica Aplicada I, Universitat Polit\`{e}cnica de Catalunya\\
Avda. Diagonal 647  08028-Barcelona. Spain. }

\begin{document}
\maketitle
\newpage
\begin{abstract}
The goal of branch length estimation in phylogenetic inference is to
estimate the divergence time between a set of sequences
based on compositional differences between them.
A number of software is currently available facilitating branch
lengths estimation for homogeneous
and stationary evolutionary models.
Homogeneity of the evolutionary process imposes fixed rates of evolution
throughout the tree. In complex data problems this assumption 
is likely to put the results of the analyses in question. 

In this work we propose an algorithm for parameter and branch lengths inference in the discrete-time Markov processes on trees. 
This broad class of nonhomogeneous models comprises the general Markov
model and all its submodels,  including both stationary and nonstationary models.
Here, we adapted the well-known Expectation-Maximization
algorithm 
and present a detailed performance study of this approach for 
a selection of nonhomogeneous evolutionary models. We conducted an
extensive performance assessment on multiple sequence alignments
simulated under a variety of settings.
We demonstrated high accuracy of the tool in parameter estimation and branch
lengths recovery, proving the method to be a valuable tool
for phylogenetic inference in real life problems.    
$\empar$ is an open-source C++ implementation of the methods
introduced in this paper and is the first
tool designed to handle nonhomogeneous data. 

\Keywords{nucleotide substitution models; branch lengths;
maximum-likelihood; expectation-maximization algorithm.}
%

\end{abstract}
\newpage
\section*{}
Assuming that an evolutionary process can be represented in a
phylogenetic tree, 
the tips of the tree are assigned operational taxonomic units (OTUs)
whose composition is known. Here, the OTUs are thought of as
the DNA sequences of either a single or distinct taxa.
Internal vertices represent ancestral sequences and inferring the branch lengths of the tree provides
information about the speciation time. 

Choice of the  evolutionary model and the method
of inference have a direct impact on the accuracy and consistency of
the results \citep{SulSwo97, Fel78, BruHal99, Pen94, HueHil93, Schwartz2010}. We assume that the
sites of a multiple sequence alignment (MSA) are
independent and identically distributed (i.i.d. hypothesis of all sites
undergoing the same process without an effect on each other),
the evolution of a set of OTUs  along a phylogenetic tree $\tree$
can be modeled by the evolution of a single character under a hidden
Markov process on $\tree$. 

Markovian evolutionary processes assign a conditional substitution (transition)
matrix to every edge of $\tree$. 
Most current software packages are based on the continuous-time Markov processes  where
the transition matrix associated to an edge $e$ is given in the form $\exp(Q^e
t_e)$, where $Q^ee$ is an instantaneous mutation rate matrix. Although
in some cases the rate matrices 
are allowed to
vary between different lineages (cf. \cite{Galtier1998},\cite{
  YY99}), it is not uncommon to equate them to a \textit{homogeneous}
rate matrix $Q$, which is constant for every lineage in $\tree$. 

Relaxing the homogeneity assumption is an important step towards
increased reliability of inference (see \cite{mitoch}).
In this work, we consider a class of processes more general than the homogeneous
ones: the discrete-time Markov processes. If $\tree$ is rooted, these
models are given by a root distribution $\pi,$ and a set of transition
matrices $A^e$ (e.g. chap. 8 of
\citet{Semple2003}). 
The transition matrices $A^e$ can freely vary for distinct edges and
 are not assumed to be of exponential form,
thus are highly applicable in the analyses of non-homogeneous data.   
Among these models we find the general Markov model ($\GMM$) and all its
submodels, e.g. discrete-time versions of the  Jukes-Cantor
model (denoted as $\JC$), Kimura two-parameters ($\ktw$) and Kimura 3-parameters
models ($\kth$), and
the strand symmetric model $\SSM$. 
Though the discrete-time models provide a more realistic fit to the data \citep{YY99,
  Ripplinger2008, Ripplinger2010}, their complexity
requires a solid inferential framework for accurate parameter estimation.
 In continuous-time models, \textit{maximum-likelihood estimation} (MLE)
was found to outperform  Bayesian methods \citep{Schwartz2010}.
The most popular programs of phylogenetic inference (PAML
\cite{Yang1997}, PHYLIP \cite{Felsenstein1989}, PAUP* \cite{PAUP}) are
restricted to the homogeneous models. 

Though more realistic, the use of nonhomogeneous models in phylogenetic inference is not yet
an established practice. 

Recently, \cite{Jayaswal2011} proposed two new non-homogeneous models.
With the objective of testing stationarity, homogeneity and 
inferring the proportion of invariable sites, the authors propose an iterative
procedure based on the \emph{Expectation Maximization}
($\ema$) algorithm to estimate parameters of the non-homogeneous models (cf. \cite{barryhartigan87}). 
The $\ema$ algorithm was formally introduced by
\cite{Dempster1977} (cf. \citet{Hartley1958}). It is a popular tool to handle
incomplete data problems or problems that can be posed as
such (e.g. missing data problems, models with latent variables, mixture
or cluster learning). This iterative procedure globally optimizes all the parameters conditional
on the estimates of the hidden data and computes the
maximum likelihood estimate in the scenarios, where, unlike in the  fully-observed model, the analytic
solution to the likelihood equations are rendered intractable.
An exhaustive list of references and
applications can be found in \cite{Tanner1996}, and more recently
in \cite{Ambroise1998}.

Here, we extend on the work of  \cite{Jayaswal2011}  and present $\empar$, a MLE method
based on the $\ema$ algorithm which allows for estimating the parameters of the
(discrete-time) Markov evolutionary models. $\empar$ is an
implementation suitable for phylogenetic trees on any number of leaves
and currently includes the following evolutionary models: $\JC$, $\ktw$, $\kth$, $\SSM$ and $\GMM.$

We test the
proposed method on simulated data and analyze the accuracy of
the parameter and branch length recovery. The tests are conducted in a
settings analogue to that of  \cite{Schwartz2010} and evaluate the
performance of $\empar$ on the 
four  and six-taxon trees with several sets of branch lengths, $\jc$
and $\kth$ models under varying alignment lengths.  We present an  in-depth
theoretical study, investigating the dependence of the performance  on factors such as
model complexity, size of the tree, positioning of the branches,
data and total tree lengths.

Our findings suggest that the method is a reliable tool for parameter
inference of small sets of taxa, best results obtained for shorter branches.

The algorithm underlying $\empar$ was implemented in C++ and is freely available to
download at \url{http://genome.crg.es/cgi-bin/phylo_mod_sel/AlgEmpar.pl}.

\section*{METHODS}

\subsection*{Models}
We fix a set of $n$ taxa labeling the leaves of a rooted
tree $\tree$. We denote by  $N(\tree)$ the set of all nodes of $\tree$, the set of leaves as $L(\tree)$,
the set of interior nodes as $Int(\tree),$ and the set of edges as
$E(\tree).$ We are given a DNA multiple sequence alignment (MSA) associated
to the taxa in $\tree$ and a discrete-time Markov process on $\tree$
associated to an evolutionary by a model $\m$, where the nodes in$\tree$ are discrete random variables with values in
the set of nucleotides $\{\a,\c,\g,\t\}$. 
We assume that all sites in the alignment are i.i.d. and model
evolution per site as follows: for each edge $e$ of $\tree$ we collect
the conditional probabilities $P(y|x,e)$ (nucleotide $x$ being
replaced by $y$ at the descendant node of $e$) in a transition matrix $A^e=(P(y|x,e))_{x,y}$;
$\pi=(\pi_{\a},\pi_{\c},\pi_{\g},\pi_{\t})$ is the distribution of
nucleotides at the root $r$ of $\tree$ and $\xi=\{\pi, (A^e)_{e}\}$
the set of  continuous parameters of  $\m$ on $\tree$. We denote by $X$ the set of $4^n$ possible patterns at the leaves
and $Y$ the set of $4^{|Int(\tree)|}$ possible patterns at the
interior nodes of $\tree.$ In what follows, the joint probability of
observing $\textbf{x}=(x_l)_{l \in L(\tree)} \in X$ at the
leaves and nucleotides $\textbf{y}=(y_v)_{v\in
Int(\tree)} \in Y$ at the interior nodes in $\tree$ is calculated as
$$p_{\textbf{x},\textbf{y}}(\xi)=\pi_{y_r}\prod_{v \in N(\tree)\setminus\{r\}
}A^{e_{an(v),v}}_{y_{an(v)},y_{v}}$$ where  $an(v)$ denotes a parent node of
node $v,$ $e_{an(v),v}$ is an edge from $an(v)$ to $v$ (note: if $v$
is a leaf, then  $y_v=x_v$).

In the \textit{complete model} the states at the interior nodes are
observed and the joint distribution is computed as above.  On the other hand, the \textit{observed
model} assumes the variables at the interior nodes
to be latent. In the latter case, the probability of observing $\textbf{x}=(x_l)_{l \in
L(\tree)}$ at the leaves of $\tree$ can be expressed as
$$p_{\textbf{x}}(\xi)=\sum_{\textbf{y}=(y_v)_{v\in
Int(\tree)} \in Y}p_{\textbf{x},\textbf{y}}(\xi).$$

Restricting the shape of the transition matrices $A^e$ leads to
different evolutionary models such as $\JC$, $\ktw$, $\kth$, $\SSM$
(see \citet{CFK},  \citet{arBook}, and
\citet{AllmanRhodes_chapter4}  for references and background on
the discrete-time models). The first three
are the discrete-time versions of the widely used continuous-time
JC69 \citep{Jukes1969}, K80 \citep{Kimura1980}  and K81
\citep{Kimura1981}  models.  
The Strand Symmetric model $\ssm$ (\citet{CS}) is a discrete-time
generalization to  the \texttt{HKY} model \citep{Hasegawa1985} with equal distribution of
the pairs of bases (\texttt{A,T}) and (\texttt{C,G}) at each node of the
tree. It reflects the double-stranded nature of DNA and was found to
be well-suited for long stretches of data \citep{Yap2004}. Lastly,  the general Markov
model $\GMM$ (\citet{Allman2003, Steel1994a}) is free of restrictions
on the entries of $A^e$,  non-stationary, and can be thought as a
non-homogeneous version of the general time reversible model (\cite{GTR}).

\subsection*{Expectation-Maximization algorithm}
An algebraic approach to the Expectation-Maximization ($\ema$) algorithm was first introduced in
\cite{Pachter2005}. In this work, we adapted this approach to the context of  phylogenetic trees.

Let $D$ denote a MSA recorded into a vector of $4^{|\leaves|}$ counts of
patterns $u_D=(u_\textbf{x})_{\textbf{x}\in X},$ where each
$u_\textbf{x}$ stands for the counts of a particular configuration of
nucleotides \textbf{x} at the leaves, observed as columns in the alignment. We are
interested in maximizing the likelihood function:
$$\mathcal{L}_{obs}(\xi ; u_D)=\prod_{\textbf{x}\in X}p_{\textbf{x}}(\xi)^{u_{\textbf{x}}}$$
(up to a constant). Let $U_{cD}=(u_{\textbf{x},\textbf{y}})_{\textbf{x}\in X,\textbf{y}\in
  Y}$ be an array of counts for the complete model, where
$u_{\textbf{x},\textbf{y}}$ is the number of times $\textbf{x}$ was
observed at the leaves and $\textbf{y}$ at the interior nodes. The
likelihood for the complete model has a multinomial form
\begin{equation}\label{eq:likelEM}
\mathcal{L}_c(\xi ; U_{cD})=\prod_{\textbf{x}\in X,\textbf{y}\in Y}p_{\textbf{x},\textbf{y}}(\xi)^{u_{\textbf{x},\textbf{y}}}=\prod_{\textbf{x}\in X,\textbf{y}\in Y}(\pi_{y_r}\prod_{v \in N(\tree)\setminus\{r\}
}A^{e_{an(v),v}}_{y_{an(v)},y_{v}})^{u_{\textbf{x},\textbf{y}}}
\end{equation}
(up to a constant), which  is guaranteed to have a global maximum given by a model-specific
explicit formula (see $\supp$ A).

$\ema$ algorithm iteratively  alternates between the expectation ($\estep$) and
maximization step ($\mstep$). In the $\estep$ the algorithm uses the tree topology, 
current estimates of parameters and the observed data $u_D$ to
assign a posterior probability to each of the possible
$4^{|\leaves|}$ patterns in $X$ and the expected counts of the
complete model, $u_{cD}$. This step can be
efficiently performed using the peeling algorithm of \cite{Felsenstein2004}.
In the $\mstep$ the updated MLE of the parameters
are obtained by maximizing the likelihood
of the complete model \eqref{eq:likelEM}. The procedure is depicted in Fig. 1. 

\begin{figure}[h!] \hspace{14mm}
\includegraphics[width=0.9\textwidth]{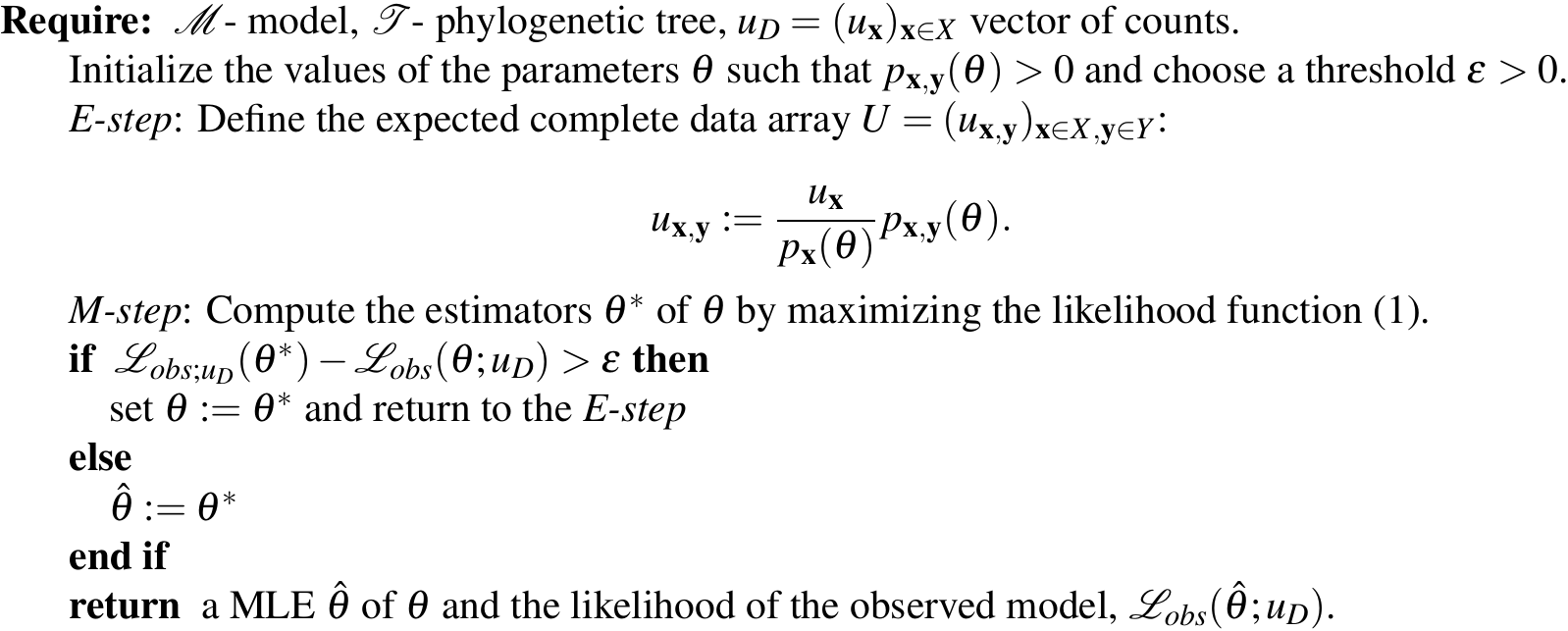} 
\caption{\small Expectation-maximization algorithm.}\label{fig:em_procedure}
\end{figure}

The likelihood is guaranteed to increase at each iteration of this process
(e.g. \cite{Wu}, \cite{Husmeier2005}). Moreover, for a compact set of parameters the algorithm converges to a critical point of the likelihood function.  Although the output of the algorithm is not guaranteed to be a global maximum, multiple starting
points are used for optimal solution.

\subsection*{Statistical tests}
The substitution matrices are assumed stochastic. The number $d$ of free parameters for transition matrices in $\JC$,
$\ktw$, $\kth$, $\SSM$ and $\GMM$ models is 1, 2, 3, 6, and 12
respectively. The root  distribution under  the $\SSM$ (respectively $\GMM$) model  has 2
(resp. 3) free parameters, and is uniform for the
remaining models considered here. For clarity of exposition, hereon
the reference to the root distribution will be omitted; however, the formulas can
be easily modified to include the root.

We let $\xi^e$ be the vector of free parameters defined as the
off-diagonal elements of a transition matrix $A$ associated to a edge
$e$ ($\xi^e_1=A_{1,2}$, $\xi^e_2$ is the next --from left to right, top down--
off-diagonal entry distinct from $\xi^e_1$, etc). The procedure is
repeated until $\xi^e_d$ is reached.


Let $\xi=(\xi_i^e)_{i=1,\ldots,d; e\in\edges}$ denote the vector of free parameters for an evolutionary model $\m$ as above
and let $\hat{\xi}$
be its MLE. Under certain regularity conditions \citep[Chap. 5]{Zacks1971},
 $\hat{\xi}$ exists, is consistent, efficient and asymptotically normal
with mean $\xi$ and the covariance matrix given by the inverse of the Fisher
information matrix \citep{Rao1973, Efron1978}.
The entries of the $d|\edges|\times
d|\edges|$ Fisher information matrix $\fish$
over free parameters are given by:
\begin{equation}
\fish(\xi^e_k, \xi^{e'}_{k'}) = -\mathbf{E}\left(\frac{\partial^2
      \log\lik_{obs}(\xi; u_D)}{\partial
      \xi^{e}_{k}\partial \xi^{e'}_{k'}}\right)
\end{equation}
(see $\supp$ B  for details). The Wald statistics for testing the null hypothesis
$\xi_i^e=\hat{\xi}_i^e$, $e\in\edges, i=1,\ldots,d$,
is
\begin{equation}\label{eq:null_distr}
(\hat{\xi^e}-\xi^e)^T \fish^e(\hat{\xi^e}-\xi^e)\sim\chi^2_{d},
\end{equation}
where $\fish^e$ denotes the $d\times d$ slice of $\fish$
corresponding to the free parameters of $e\in\edges$. The $p-value$ can
thus be easily calculated by looking at the tails of the
corresponding $\chi^2$ distribution. 

We tested the validity of the test statistics in our data by simulating a
variety of MSAs under the complete model and compared it to the
theoretical distribution \eqref{eq:null_distr}. Figure I in the
$\supp$ C shows high fit and proves that the
setting is appropriate.

Variances of the free parameters of the model and the full (observed) covariance matrix 
are saved in the output of $\empar$. These in turn can be used as the plug-in
estimators in \eqref{eq:null_distr} to calculate the $p-values$ and normal confidence intervals for the parameters.

We denote by $V^{e}_{i,i}$ the $i^{th}$ diagonal entry of the matrix
$(\fish^e)^{-1}$ corresponding to the variance of the free
parameter $\xi^e_i$, $i=1,\ldots,d$. For the models with $d>1$ (i.e. all
but $\JC$), the variances of 
the free parameters can be summarized in a combined form $cV^e$ for each edge $e$:
\begin{equation}\label{eq:comb_var}
cV^{e}(\xi^e)=\frac{\sum^d_{j=1} \left(V^e_{j,j}+\left(\xi^e_j-\frac{\sum^d_{j=1}\xi^e_j}{d}\right)^2\right)}{d}.
\end{equation}

\subsection*{Branch lengths}
The evolutionary distance between two nodes in $\tree$
joined by an edge $e$ with substitution matrix $A^e$ is defined as the
total number of substitutions per site along $e$. This quantity is
referred to as the \textit{branch length} of edge $e$ (or of
matrix $A^e$) and, following on
\citep{barryhartigan87trans}, can be approximated by:
\begin{equation}
l(\me)=-\frac{1}{4}\log\det(\me).\label{eq:brlength}
\end{equation}
We denote the \textit{total length of the tree} $\tree$ by $\len$,
$\len=\sum_{e\in|\edges|}l(\me).$

Now, let $A$ and $A'$ be two invertible $4\times 4$ matrices such that
the entries of $(A')^{-1}(A-A')$ are small. From \eqref{eq:brlength}, we get
\begin{eqnarray}\label{eq:br_bound}
|l(A)-l(A')|&=&\frac{1}{4}|\log\frac{\det(A)}{\det(A')}|=\frac{1}{4}|\log\det((A')^{-1}A)|
\nonumber\\
&=&\frac{1}{4}|\log\det(\textbf{Id} +
(A')^{-1}(A-A'))|\nonumber\\&\approx& \frac{1}{4}|\log(1+Tr((A')^{-1}(A-A')))|\nonumber\\
&\approx&\frac{1}{4}|Tr((A')^{-1}(A-A'))|\leq \frac{1}{4}4||(A')^{-1}(A'-A))||_{1} \nonumber\\
&\leq& ||(A')^{-1} ||_{1} ||A-A'||_{1},
\end{eqnarray}
where $||.||_1$ is the maximum absolute column sum of the
matrix. Therefore if $A'$ is a good approximation to $A$, then $l(A')$
is a good approximation to $l(A)$. In what follows, we use the statistical test above to
show the accurate recovery of the parameters. By the above argument,
we can conclude that the estimates of the branch lengths will also be accurate (see also Results section).

\section*{Simulated data}
Performance assessment of $\empar$ was conducted on the MSAs simulated
on four and six-taxon trees following \citep{Schwartz2010}. In the case of four
taxon trees we fixed an inner node as the root and considered three types of topologies: $\sone^4$ corresponds to
the ``balanced'' trees with all five branches equal; the inner branch
in $\stwo$ is half the length of the exterior branches;
and $\sth$ denotes a topology with the inner
branch double the length of the external ones (see
Fig.~2). 
In $\sone^4$ and $\sth$ we let the length $l_0$ of the inner branch vary from
0.01 to 1.4, where starting from 0.05 it increases in steps of
0.05; in $\stwo$ we let $l_0$ vary in $(0,0.7).$  For 6-taxon
trees we used only balanced trees $\sone^6$ (see Fig.~2)
with $l\in (0,0.7)$.

\begin{figure}[h!] \hspace{2mm}
\includegraphics[width=0.2\textwidth]{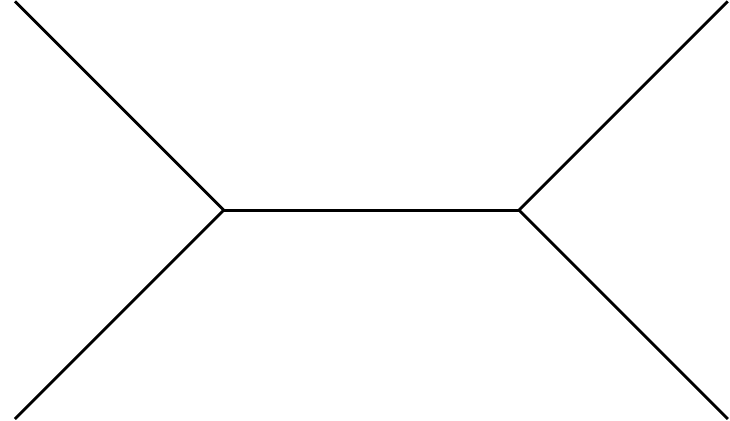} \hspace{2mm}
\includegraphics[width=0.2\textwidth]{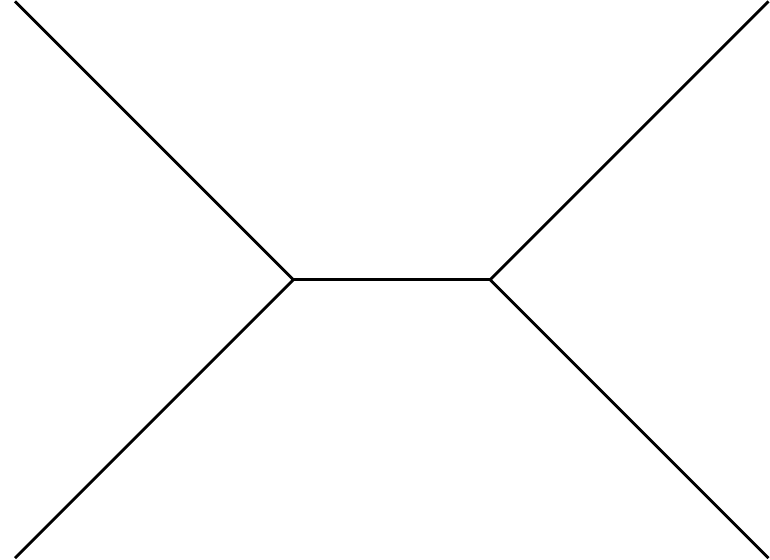}\hspace{2mm}
\includegraphics[width=0.2\textwidth]{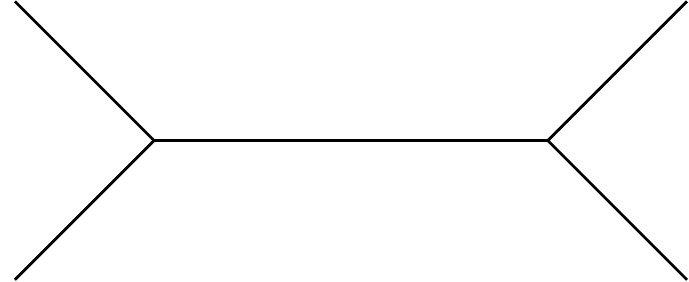}
\hspace{5mm}
\includegraphics[width=0.2\textwidth]{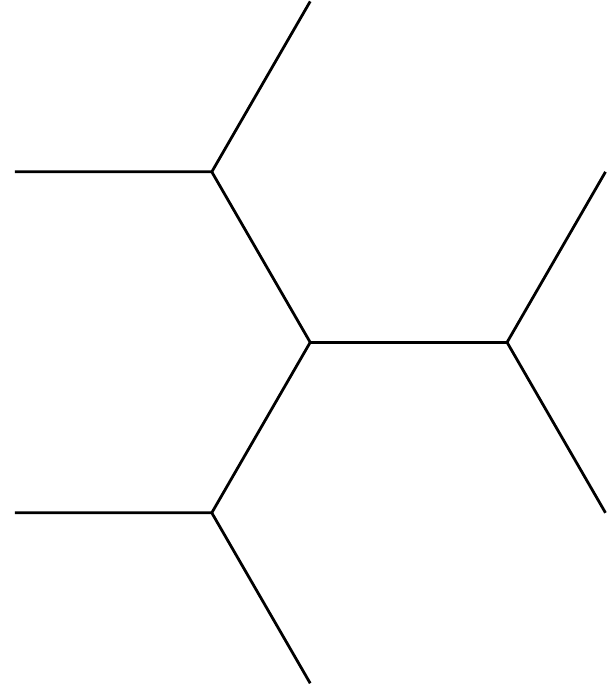}
\caption{\small Unrooted trees used for simulations: $\sone^4$, $\stwo$, $\sth$ and $\sone^6$
  (\emph{from left to right}).}\label{fig_trees}
\end{figure}

We simulated multiple sequence alignments on trees with 4 and 6
leaves under $\JC$ and $\Kb$ models. We used the
$\texttt{GenNon-H}$ package available from \url{http://genome.crg.es/cgi-bin/phylo_mod_sel/AlgGenNonH.pl}.
In brief, based on an input phylogenetic tree with given branch lengths, $\texttt{GenNon-H}$ samples the substitution
matrices corresponding to these lengths for all edges and
uses them to generate the DNA MSAs following discrete-time Markov process on the tree. The output of this software is the
alignment, the substitution matrices, root distribution (whenever non-stationary) and the
variances of the continuous free parameters.
We note that for the $\jc$ model, there is a $1-1$ correspondence between the
branch length and the free parameters of the substitution matrix. This does not hold for other models, were different substiution matrices may give the same branch length.
%

We set the alignment  length $L$ to $300nt, 500nt, 1,000nt$ and $10,000nt$
for 4-taxa and to $1,000nt$ or $10,000nt$ for 6-taxa.  For the
$\JC$ and $\Kb$ evolutionary models, a phylogenetic tree $\tree$ (with
branch lengths), and a given alignment length, we run each analysis
$1,000$ times. 
and estimated the parameters using $\empar$. 


All MSAs used for the tests are accessible at
the $\empar$ webpage.

\subsection*{Identifiability}
 
It is known that in certain cases the substitution parameters are not identifiable
(e.g. parameters at the edges adjacent to the root of valency 2).
As shown in \cite{Allman2003}, the $\GMM$ model and its
submodels, are identifiable up to a permutation of rows. 
\cite{BHident} showed that incorrect order of rows in the matrices
can lead to a negative determinant of the substitution matrix from which the branch lengths cannot be calculated.

We expected this problem to arise in short data sets and large branch length, as those
correspond to the substitution matrices with smaller diagonal value.
For all the data sets used for tests, we calculated the percentage of cases among the 1,000 simulations
for which the parameters estimated by the EM algorithm were
permuted. This phenomenon was only observed in the data sets of $300nt$ and
$1,000nt$. In the first case, 
the estimated matrices were permuted when the initial branch length
was 0.55 or longer and 
corresponded to 0.005-0.023$\%$ of the cases; in the latter, for the
branches of 0.6 or longer with at most 
0.001$\%$ permuted matrices. Shorter branch lengths and longer
alignments did not suffer from the above problem 
and recovered the underlying order in all of the cases.

As shown by \cite{Chang96}, the entries of the Diagonal Largest in
Column (DLC) substitution matrices are identifiable.
Namely,  there exists a unique set of substitution matrices satisfying
the DLC condition and a unique root distribution that leads to a given
joint distribution at the leaves. 
In order to ensure the reliability of the results we designed a
procedure that scans the tree in the search of the permutations that
maximize the number of substitution matrices with larger diagonal
entries. It is not possible to maximize it for all edges, thus the goal is to find the permutations giving more weights to the lower parts of the tree, starting with the nodes corresponding to the outer branches.
Given a tree $\tree$, we choose an interior node to be
the root, directing all edges outwards. For each interior node
$x$, we apply a permutation $S(x)$ of $\{\a,\c,\g,\t\}$ that maximizes
the sum of diagonal entries of the matrices assigned to the outgoing
edges of $x$. 
Permutations $S(x)$ are applied recursively to the subtrees of
$\tree$, moving $x$ from the outer nodes towards the root.


\section*{RESULTS AND DISCUSSION}
We present the results on the simulated data sets and discuss
their dependence on the length of the alignments, the length of
the branches and the depth of the branches in the tree--1 for the external branches and 2 for
the internal branches \citep{Schwartz2010}. In cases with multiple
branches of equal depth, we chose one of them at random. 

\subsection*{Results of statistical tests}
Each sample gave rise to
a $p-value$ based on the $\chi^2_d$ test given by \eqref{eq:null_distr}.
The $p-values$ are a measure of strength of evidence against the null
hypothesis: for both exceptionally small or large $p-values$ one can reject the null hypothesis.

We recorded the proportion of samples for which the
$p-value$ lied in the interval $(0.05,0.95)$. The results are shown Table~\ref{tab:pval_jc_short} for the $\jc$
model on the $\stwo$ tree (also see Tab. I-V in the $\supp$ C). 
We observe that even for short alignments of 300nt the null hypothesis cannot be rejected in approximately 95\% of the tests. 

\begin{table}[h!]
\caption{\small The relative frequency of \emph{p-value}
  $\in(0.05,0.95)$  among the 1,000 $\chi^2$ tests based on the asymptotic normality of
  the maximum likelihood estimator under the $\jc$ model on the
  $\stwo$ tree. 
The first column indicates the lengths of the depth 2 branch of $\stwo$. 
The results are presented for both depths of the branches.
  }
\centering
  \tiny
  \begin{tabular}{p{0.8cm}p{0.8cm}p{0.8cm}p{0.8cm}p{0.8cm}|p{0.8cm}p{0.8cm}p{0.8cm}p{0.8cm}}
& \multicolumn{4}{c|}{depth 1} & \multicolumn{4}{c}{depth 2} \\
$\mathbf{l}$ $\mid$ L & \multicolumn{1}{r}{300nt} &
\multicolumn{1}{r}{500nt} & \multicolumn{1}{r}{1,000nt} &
\multicolumn{1}{r|}{10,000nt} & \multicolumn{1}{r}{300nt} &
\multicolumn{1}{r}{500nt} & \multicolumn{1}{r}{1,000nt} &
\multicolumn{1}{r}{10,000nt} \\\cline{2-9}
  0.01 & 0.971 & 0.972 & 0.968 & 0.946 & 0.972 & 0.949 & 0.868 & 0.958 \\ \cline{2-9}
    0.05 & 0.947 & 0.951 & 0.947 & 0.948 & 0.974 & 0.943 & 0.953 & 0.952 \\ \cline{2-9}
    0.10 & 0.949 & 0.953 & 0.964 & 0.952 & 0.952 & 0.948 & 0.948 & 0.955 \\ \cline{2-9}
    0.15 & 0.952 & 0.954 & 0.958 & 0.938 & 0.946 & 0.953 & 0.940 & 0.947 \\ \cline{2-9}
    0.20 & 0.957 & 0.944 & 0.944 & 0.954 & 0.949 & 0.965 & 0.944 & 0.954 \\ \cline{2-9}
    0.25 & 0.957 & 0.955 & 0.955 & 0.956 & 0.945 & 0.939 & 0.955 & 0.936 \\ \cline{2-9}
    0.30 & 0.957 & 0.943 & 0.945 & 0.955 & 0.943 & 0.946 & 0.941 & 0.948 \\ \cline{2-9}
    0.35 & 0.952 & 0.943 & 0.958 & 0.958 & 0.948 & 0.943 & 0.950 & 0.960 \\ \cline{2-9}
    0.40 & 0.955 & 0.946 & 0.947 & 0.957 & 0.951 & 0.951 & 0.936 & 0.944 \\ \cline{2-9}
    0.45 & 0.949 & 0.944 & 0.944 & 0.947 & 0.948 & 0.955 & 0.958 & 0.958 \\ \cline{2-9}
    0.50 & 0.948 & 0.935 & 0.942 & 0.941 & 0.929 & 0.949 & 0.954 & 0.946 \\ \cline{2-9}
    0.55 & 0.954 & 0.949 & 0.946 & 0.957 & 0.936 & 0.944 & 0.944 & 0.952 \\ \cline{2-9}
    0.60 & 0.940 & 0.942 & 0.937 & 0.953 & 0.944 & 0.934 & 0.948 & 0.955 \\ \cline{2-9}
    0.65 & 0.940 & 0.934 & 0.955 & 0.952 & 0.938 & 0.938 & 0.945 & 0.948 \\ \cline{2-9}
    0.70 & 0.944 & 0.936 & 0.942 & 0.946 & 0.917 & 0.940 & 0.944 & 0.948 \\ \cline{2-9}
    0.75 & 0.922 & 0.932 & 0.947 & 0.934 & 0.922 & 0.932 & 0.943 & 0.950 \\ \cline{2-9}
    0.80 & 0.909 & 0.932 & 0.926 & 0.957 & 0.957 & 0.928 & 0.943 & 0.941 \\ \cline{2-9}
    0.85 & 0.912 & 0.912 & 0.932 & 0.948 & 0.968 & 0.930 & 0.936 & 0.947 \\ \cline{2-9}
    0.90 & 0.870 & 0.885 & 0.919 & 0.951 & 0.980 & 0.918 & 0.929 & 0.953 \\ \cline{2-9}
    0.95 & 0.852 & 0.888 & 0.939 & 0.951 & 0.981 & 0.965 & 0.908 & 0.944 \\ \cline{2-9}
    1,00 & 0.824 & 0.866 & 0.893 & 0.935 & 0.982 & 0.981 & 0.896 & 0.933 \\ \cline{2-9}
    1,05 & 0.816 & 0.853 & 0.889 & 0.930 & 0.980 & 0.981 & 0.898 & 0.937 \\ \cline{2-9}
    1.10 & 0.806 & 0.852 & 0.891 & 0.921 & 0.990 & 0.995 & 0.925 & 0.945 \\ \cline{2-9}
    1.15 & 0.784 & 0.812 & 0.867 & 0.938 & 0.980 & 0.987 & 0.982 & 0.951 \\ \cline{2-9}
    1.20 & 0.797 & 0.785 & 0.823 & 0.923 & 0.986 & 0.986 & 0.984 & 0.942 \\ \cline{2-9}
    1.25 & 0.786 & 0.803 & 0.824 & 0.938 & 0.983 & 0.981 & 0.984 & 0.941 \\ \cline{2-9}
    1.30 & 0.789 & 0.793 & 0.800 & 0.894 & 0.981 & 0.976 & 0.992 & 0.925 \\ \cline{2-9}
    1.35 & 0.755 & 0.787 & 0.786 & 0.893 & 0.973 & 0.991 & 0.989 & 0.912 \\ \cline{2-9}
    1.40 & 0.761 & 0.789 & 0.785 & 0.864 & 0.970 & 0.974 & 0.994 & 0.879 \\ \cline{2-9}
  \end{tabular}\label{tab:pval_jc_short}
\end{table}

\subsection*{Error in transition matrices}
For a given branch, we quantified the divergence $D$ between the
original and estimated parameters of its transition matrix $A$ using the
induced $L_1$ norm: $||A-\hat{A}||_{1}$ (see \eqref{eq:br_bound}).
The columns in the transition matrices of $\jc$, $\mathtt{K80}^{\ast}$, and the $\kth$ are
equal and the norm becomes:
\begin{equation}\label{eq:L2norm}
D=\sum_{i=1}^4\mid A_{i,1}-\hat{A}_{i,1}\mid.
\end{equation}
Figure 3 
depicts the results for $\JC$ and $\Kb$ on the three 4-taxon
phylogenies, different alignment lengths and depths of the branches. 
The shapes of the distribution of $D$ for both models 
are very similar.
As expected, the performance is weaker for long branches and short alignments. 
A great improvement is observed with the increase in the alignment
length, e.g.  $10,000nt$ depicts very accurate estimates.
The performance under the $\JC$ model (Fig.~3a )
is better 
than that of $\kth$ (Fig.3b) 
 for shorter branch lengths.

\subsection*{Parameter dispersion}
Figure 4 
shows the variances of the estimated
parameters for depth 1 and 2 branches on the $\sone$, $\stwo$,
$\sth$ trees under the $\jc$ model. 

The variances show an exponential increase, which is most significant
in the $\sone^4$ tree for both depths, and the $\stwo$ for depth 2. The results for the depth 1 branch in $\sone$ and $\stwo$ are
very similar. The smallest variance was observed for the depth 2 of $\sth$.  
For alignments
of length $10,000nt$ on four taxa we can say that the method is quite
accurate (see also Tab.VI-VIII in the $\supp$ C). 

For the $\kth$ model we summarized the results on variances for
each edge as the mean of combined variances of all samples (see formula \eqref{eq:comb_var}).
The results are analogous to those of the $\jc$ model, see Figure II in
$\supp$ C. As expected, the parameter estimates
are less dispersed for shorter branches and longer
alignments (see Tab. IX-XI in the $\supp$ C).

\subsection*{Error in the branch lengths}
Using the formula~\eqref{eq:brlength} we calculated the actual
difference $l_0-\hat{l}$ between the branch length $l_0$ computed from the
original parameters $\xi$ and the branch length $\hat{l}$ computed
using their MLEs $\hat{\xi^e}$.  Negative values of this score
imply overestimation of the branch length, while positive values
indicate underestimation. The results are shown in
Figures~5 and 6. 

In the case of $\jc$ we observe that the method presented here does not tend to underestimate or
overestimate the lengths for the depth 1 branches in all the 4-taxon
trees ($l_0-\hat{l}$ is centered at 0 (see Fig.~5). 
The depth 2 branches have a tendency towards overestimation of the length
for branches longer than (approximately) $0.45$ for $\stwo$, $0.9$ for
$\sth$, and  $0.8$ for the $\sone^4$ trees. In the latter case,
lengths longer than $1.2$ for alignments up to $1,000$nt show
opposite trend of underestimating the true lengths. The values were
accurate when the alignment lengths were increased in the case of
$\stwo$ and $\sth$. On the other hand, for $\sone^4$ the alignments of
$10,000$nt resulted in overestimation.


In the $\kth$ model the results are significantly more accurate (see
Fig.~ 6). 
There is a trend of underestimation for branches longer than
(approximately) $0.9$ for shorter alignments. That is especially
noticeable for $\sone^4$ and depth 1 branches of $\stwo$. This trend diminishes with an increase in
the alignment length. Overall, in the case of both models, the variance of the estimate is smaller for shorter lengths and
both depth 1 and 2 branches of the $\sth$ tree.

In addition, we calculated  the tree length $\len$ (i.e. the sum of its branch
lengths) from the estimated parameters and compared it to the
theoretical result on the original branch length $l_0$: $4.5l_0$
for $\stwo$ (where $l_0$ is a depth 1 branch), $3l_0$ for $\sth$ ($l_0$ for
depth 2 branch) and $5l_0$ for $\sone^4$. The rightmost columns of
Figures~5 and 6 
show the results for
4-taxon trees for the $\jc$ and $\kth$ models respectively.
The length of the tree is estimated accurately for all trees, the
estimates being best for $\sth$. The variance is small and decreasing with an
increase in the data length. As the sequences get longer, the
distribution is centered around the true value. This is especially
visible for the $\kth$ model (see Fig.~ 6). 

\subsection*{Results for larger trees} 
We run the analysis on the 6-taxon balanced
tree, $\sone^6$, under the $\kth$ model, for alignment lengths of $1,000$nt
and $10,000$nt  and branch length $\br\in\{0.01, 0.1, 0.3,
0.5, 0.7, 0.9, 1.1, 1.3, 1.4\}$.
The $p-values$ of the corresponding tests confirm that the performance
of the algorithm is very satisfactory (see Tab. XII in the $\supp$ C).
We have seen in the 4-taxon study that the tree with equal branch
lengths gave worse results than the unbalanced trees. Thus, we
expect the results of the depth 2 branches to be similarly challenged in
this case.

Figure~7 
depicts the estimated tree lengths. It can be
seen that the estimates are accurate and the results improve for
the alignments of $10,000$nt. As expected, the variance of the
estimates increases with the increase in the length of the branch.
By formula \eqref{eq:brlength}, long branches correspond to small values
of the determinant of the transition matrix. Thus, statistical fluctuations in the parameter
estimates have greater impact on the resulting length of the tree.

Next, we calculated the difference between
the original and estimated branch lengths. In Figure~8a 
we see that the depth 1 branches show some degree of underestimation of
the length for lengths $1.1-1.4$  and alignments of $1,000$nt.
In the case of $10,000$nt, the results improve and can be expected to
show little bias for even longer data sets. Branches of depth 2 show higher
degree of underestimation with improvement for longer data sets. 
The divergence of the original and estimated parameters for transition matrices given by formula \eqref{eq:L2norm}
is shown in Figure~8b. 
For branches of depth 1 and
data of length $10,000$, the error is about 0.2. In the case
of branches of depth 2, it is almost doubled for both alignment
lengths. In both cases, branch lengths up to 0.5 give satisfactory
results. The error of the estimates for longer branches seems to be approaching a
plateau.

Combined variance of the estimated parameters is much decreased for
the $10,000$nt data sets in comparison with the $1,000$nt, and is
smaller for the depth 1 branch (see Fig.~8c ). 
 Again, the
exponential shape of the plot can be attributed to the logarithm appearing in the formula \eqref{eq:brlength}.

\section*{CONCLUSION}

In order to evaluate the performance of the method proposed here
under various circumstances, we conducted many tests on simulated
data sets. We observed that the performance of $\empar$ is most optimal for long
alignments and short branch lengths. 

It is worth noting that even for short alignments of 300nt or 500nt
on 4 taxa, the estimated parameters approximate closely the original
parameters  in $\approx 95\%$ of the cases as proved by the normality
test of the MLE. 
Moreover, the branch lengths calculated based on the parameters
estimated by $\empar$ were found very accurate already for short
alignments. Though the measure of divergence $D$ for the parameters of transition matrices
proposed here accumulates all errors in the entries of the transition matrix, alignment length of 10,000nt showed
divergence values  smaller than 0.1.


In this paper, we provide the first implementation of a tool for inferring
continuous parameters under the discrete-time models. The method
allows for accurate estimation of branch lengths in non-homogeneous data.
There are two limitations to applicability of  the method. Firstly,
the algorithm has an exponential computational time increase 
 with the number of taxa. This
is a restriction due to the fact that the algorithm
computes large matrices of dimension that is exponential in the total
number of nodes of a tree. Running time of $\empar$ on star trees with 3-8 nodes and equal
branches of $0.5$ on Ubuntu 11.10, Intel Core i7 920 at 2.67 GHz
with 6 Gb is given in Table~\ref{tab:gennong_times}.
Secondly, the memory usage of $\empar$ is approx. $8*4^{|N(\tree)|}$
and corresponds to the memory footprint of the matrix in the $\ema$
 algorithm, e.g. for this matrix to fit
in the memory of a 6Gb machine the bound 
on the number of nodes is $|N(\tree)|$ $\leq 14$. 

\begin{table}[h!]
\centering\caption{$\empar$ performance time-- estimating the
  parameters of $\kth$ on star trees with equal branch lengths of
  $0.5$, varying number of leaves, $L(\tree)$, for the MSAs of $1,000$ and $10,000nt$.}
 \begin{tabular}{p{2.5cm}p{1.2cm}p{1.2cm}p{1.2cm}p{1.2cm}p{1.2cm}p{1.2cm}}
 length $\mid$ n & 3 & 4 & 5 & 6 & 7 & 8 \\\hline
1,000  & 0.004 & 0.02 & 0.033 & 0.222 & 1,049 & 7.14 \\\hline
10,000  & 0 & 0.011 & 0.043 & 0.171 & 1,044 & 6.95
\end{tabular}\label{tab:gennong_times}
\end{table}

We conclude that $\empar$ is a highly reliable method for estimating
branch lengths of relatively small number of taxa and trees with short branch
lengths (e.g. closely related species), and achieves high accuracy
even when the results are based on short sequences. In particular, $\empar$ is a reliable method to compute quartets and to be used with quartet-based methods (see \cite{Berry99} and \cite{Berry2000}) on  nonhomogeneous data.

\section*{FUNDING}
  \ifthenelse{\boolean{publ}}{\small}{}

Both authors were partially supported by Generalitat de Catalunya, 2009 SGR 1284. MC is partially supported by Ministerio de Educaci\'on y Ciencia MTM2009-14163-C02-02.We thank Roderic Guig\'o for generously providing funding for this project under grant BIO2011-26205 from the Ministerio de Educaci\'{o}n y Ciencia (Spain).

\newpage
\bibliographystyle{plainnat}




\listoffigures
\begin{figure}[h!]
\centering
\subfigure[$\jc$]{
 \begin{minipage}[b]{1.1\linewidth}
\includegraphics[scale=0.18]{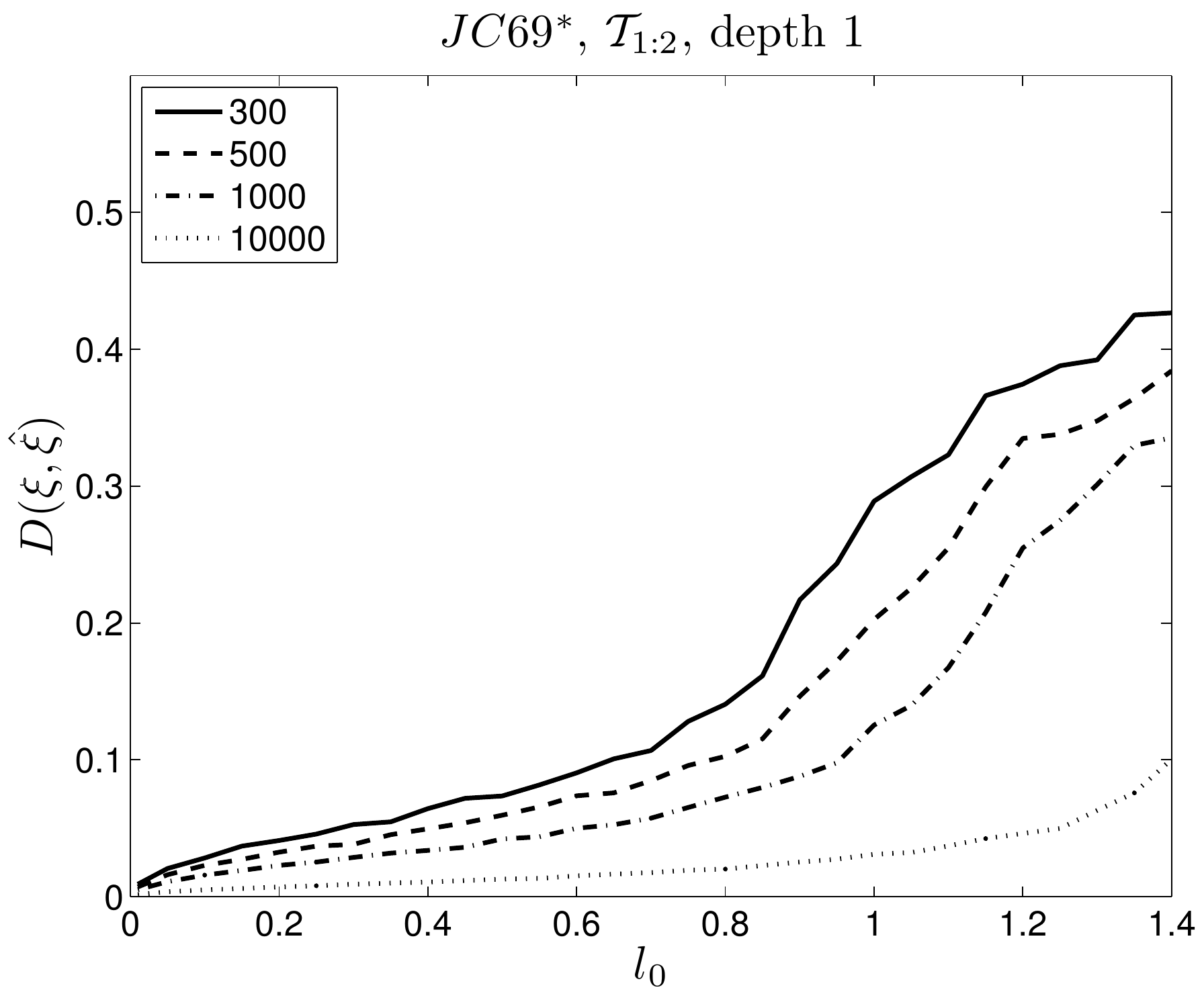}
\includegraphics[scale=0.18]{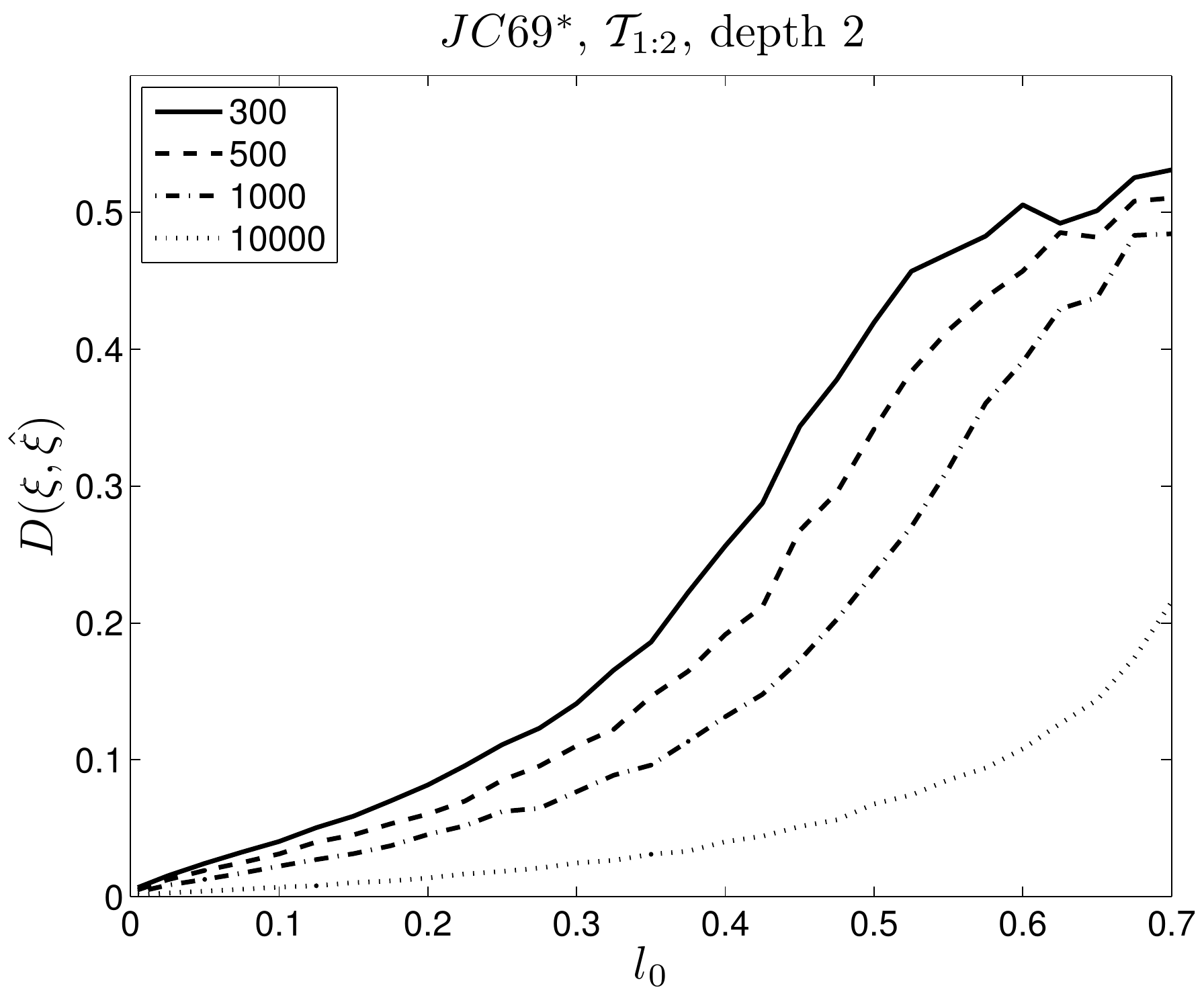}
\includegraphics[scale=0.18]{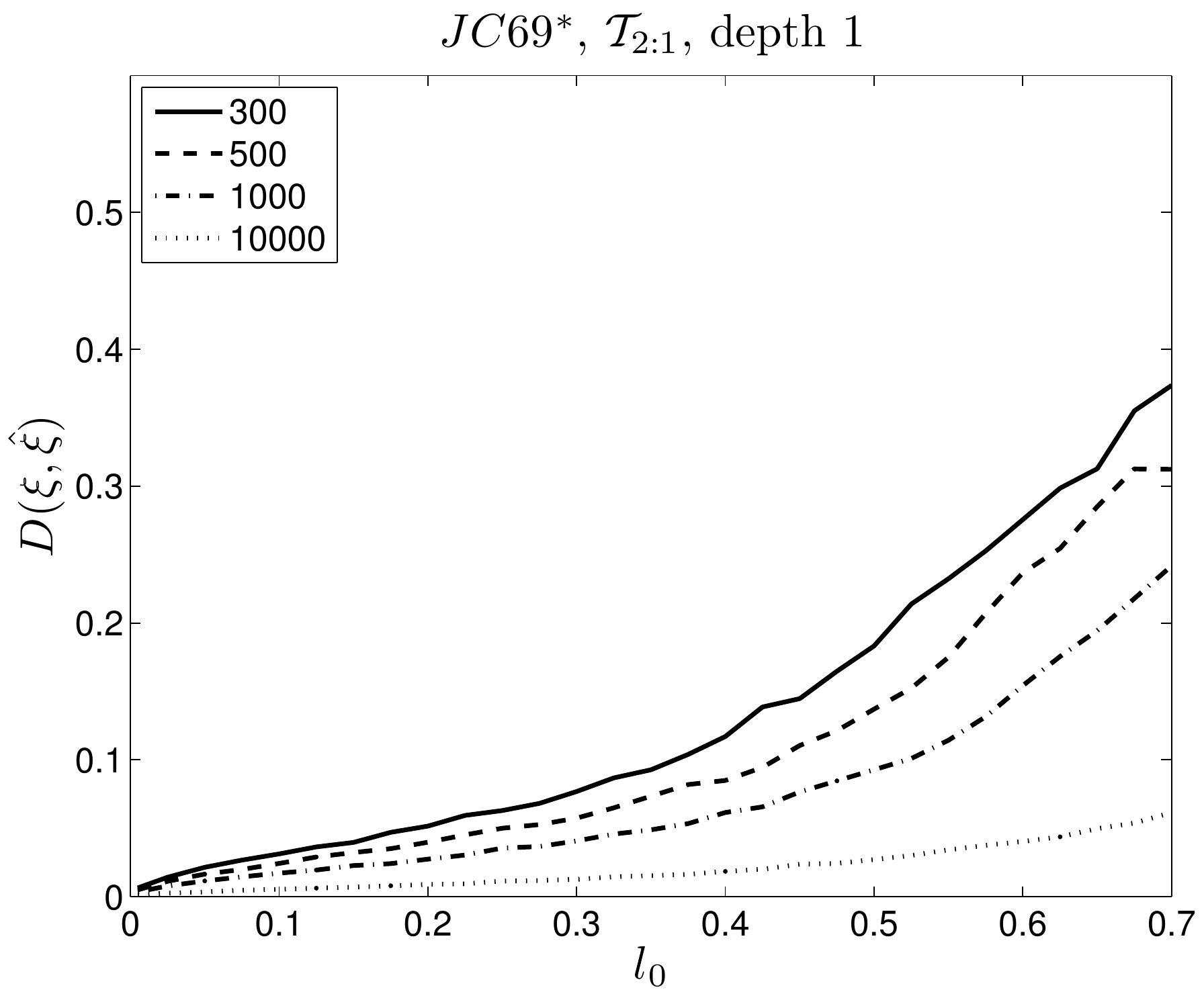}
\includegraphics[scale=0.18]{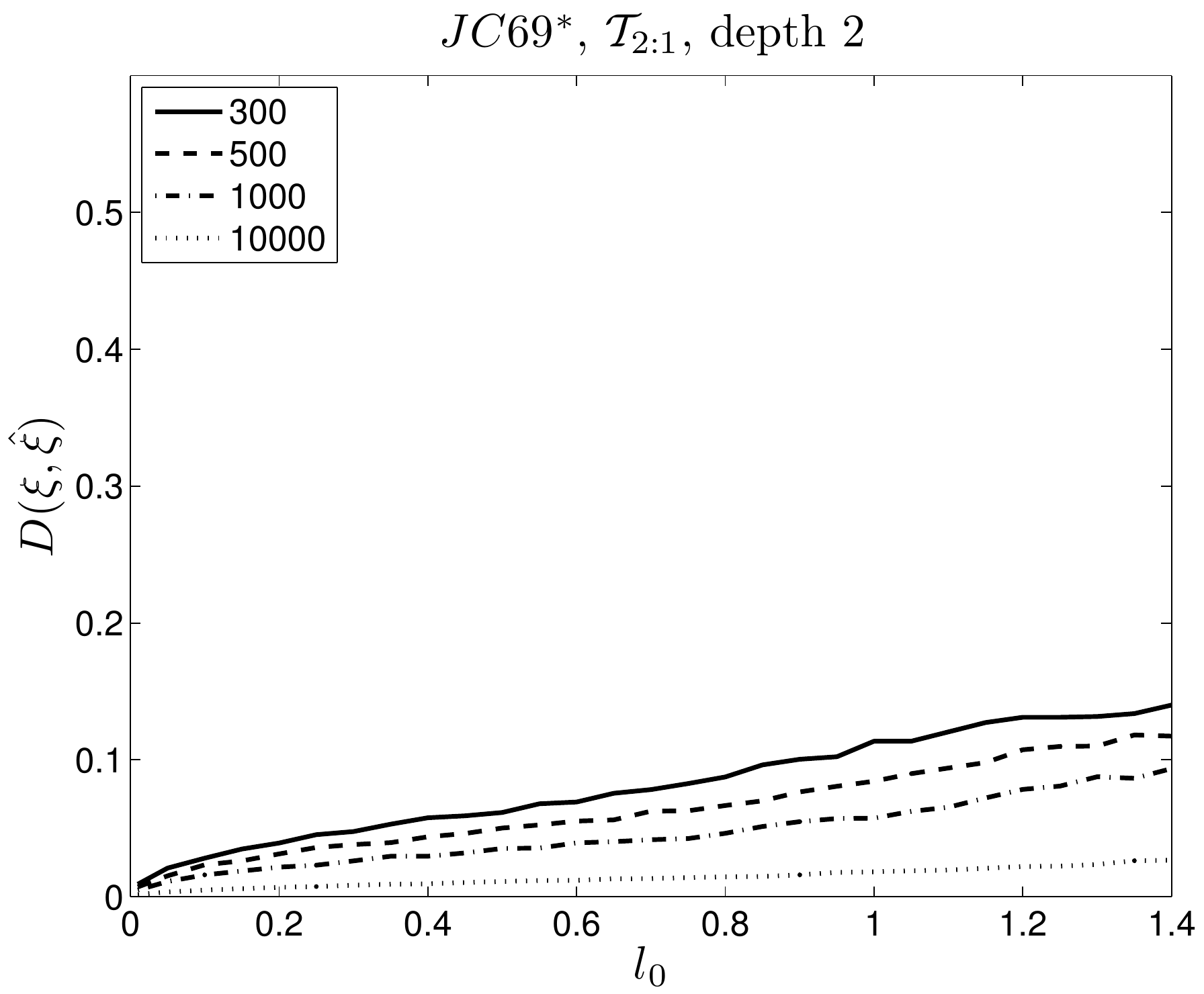}\vfill
\includegraphics[scale=0.18]{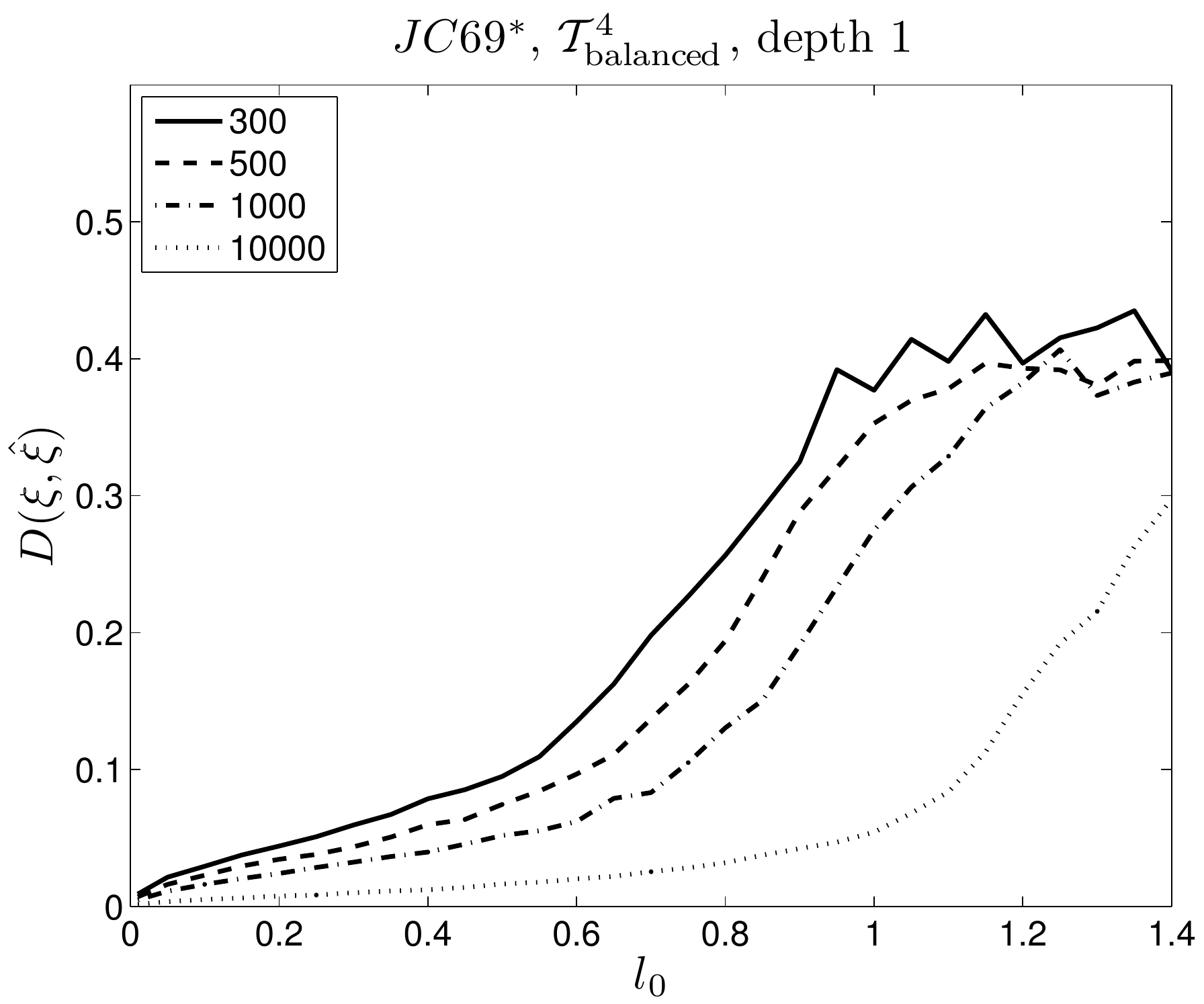}
\includegraphics[scale=0.18]{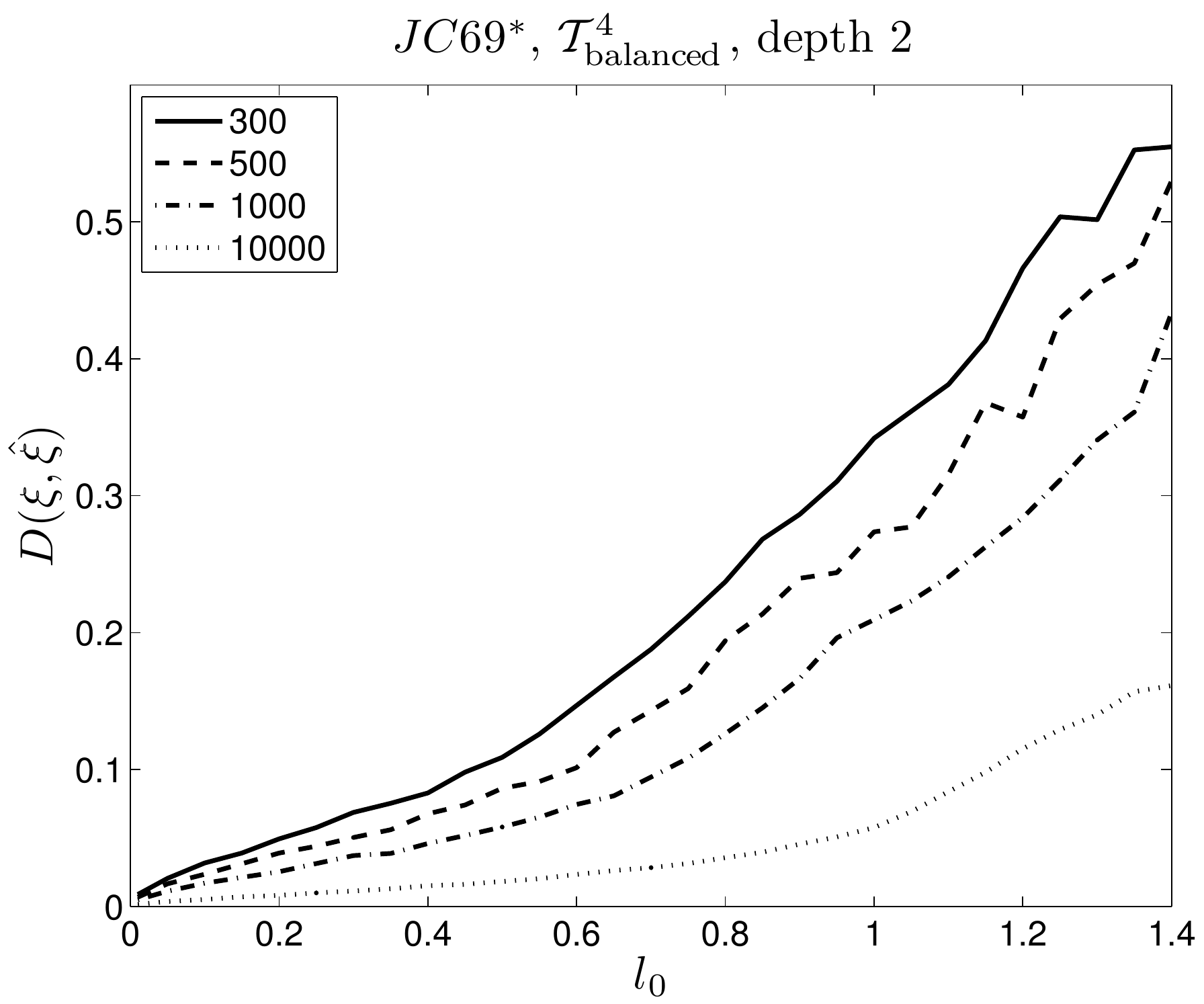}
\end{minipage}
\label{fig:jc_dist}}   
\subfigure[$\kth$]{
 \begin{minipage}[b]{1.1\linewidth}
\includegraphics[scale=0.18]{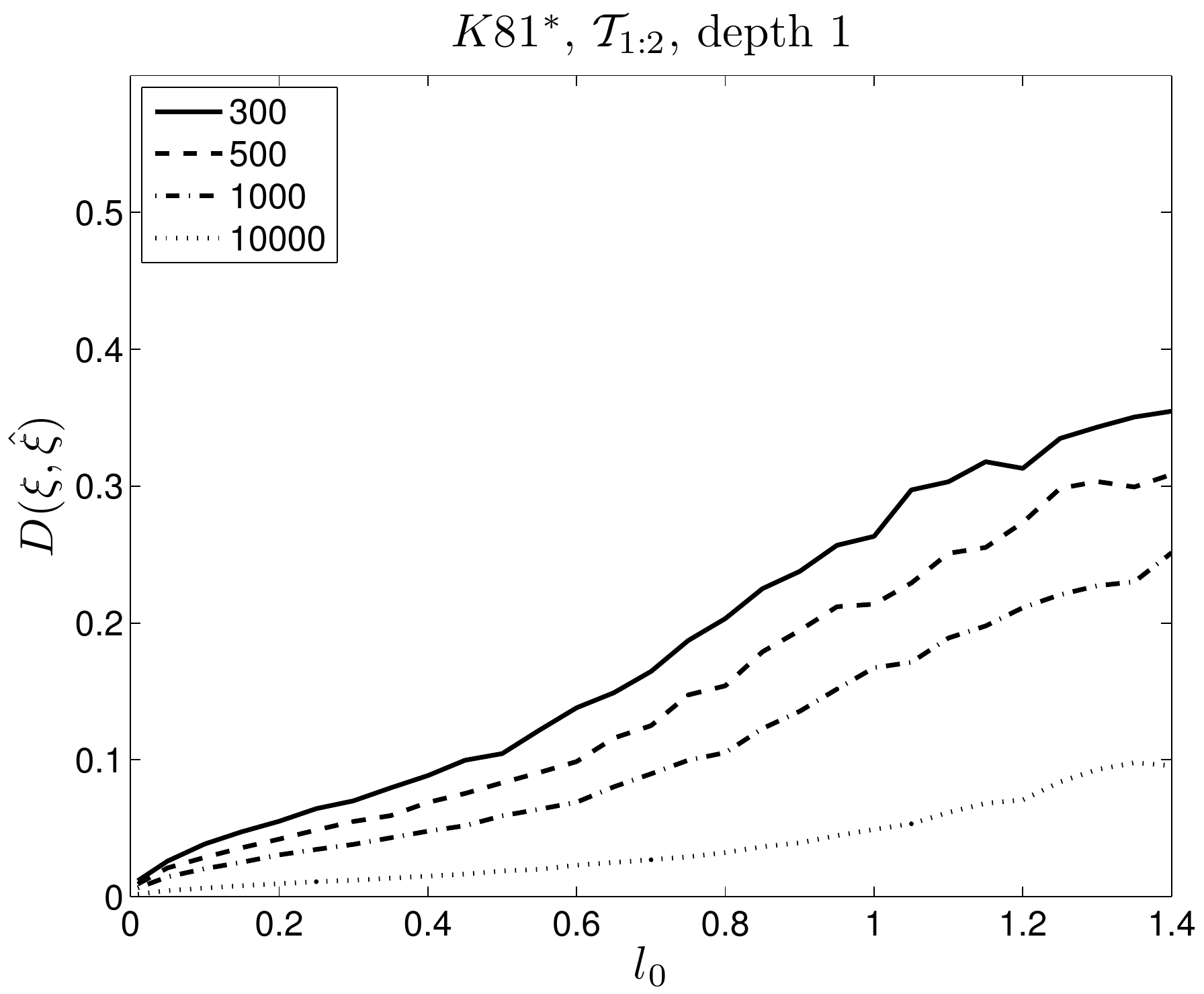}
\includegraphics[scale=0.18]{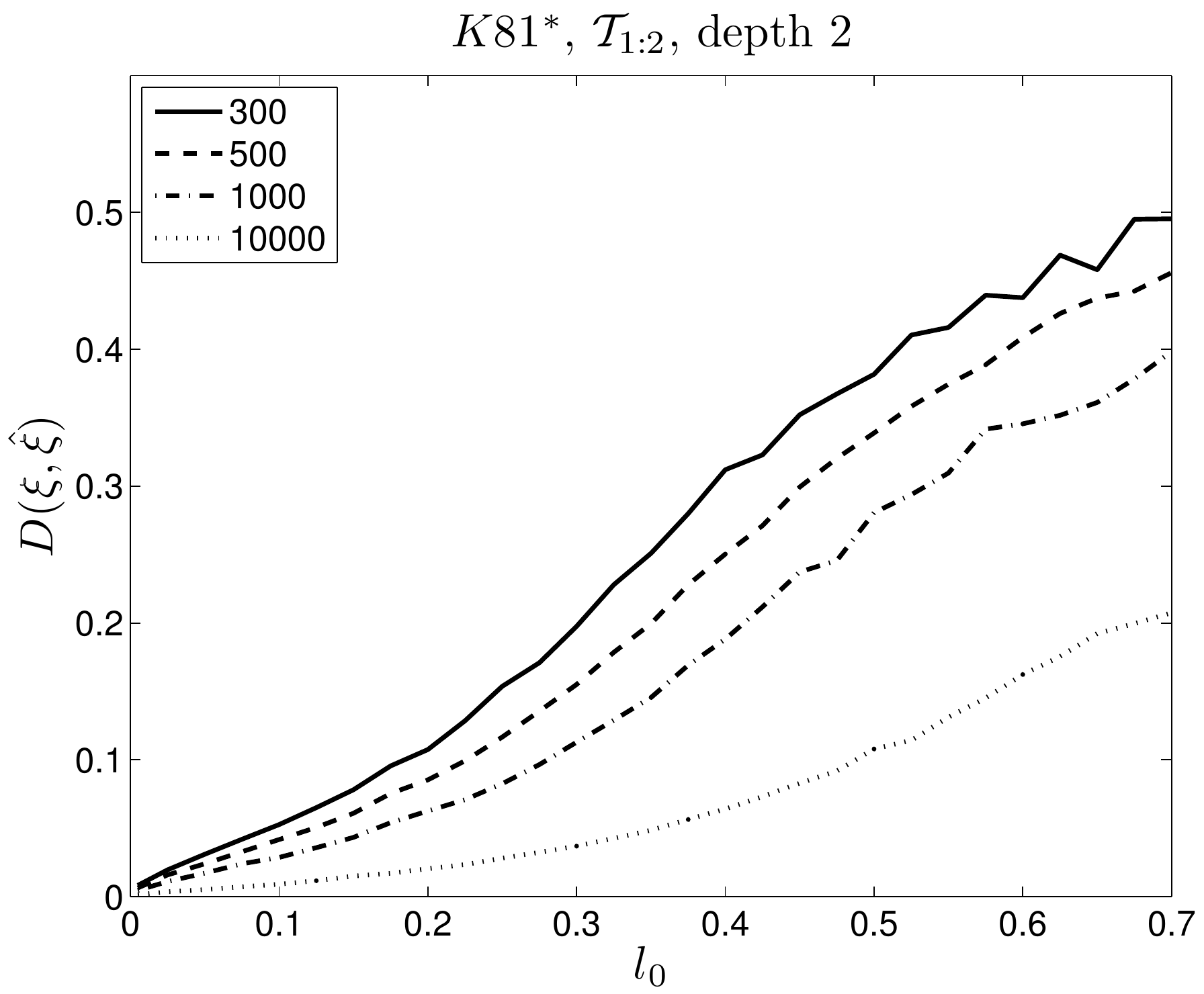}
\includegraphics[scale=0.18]{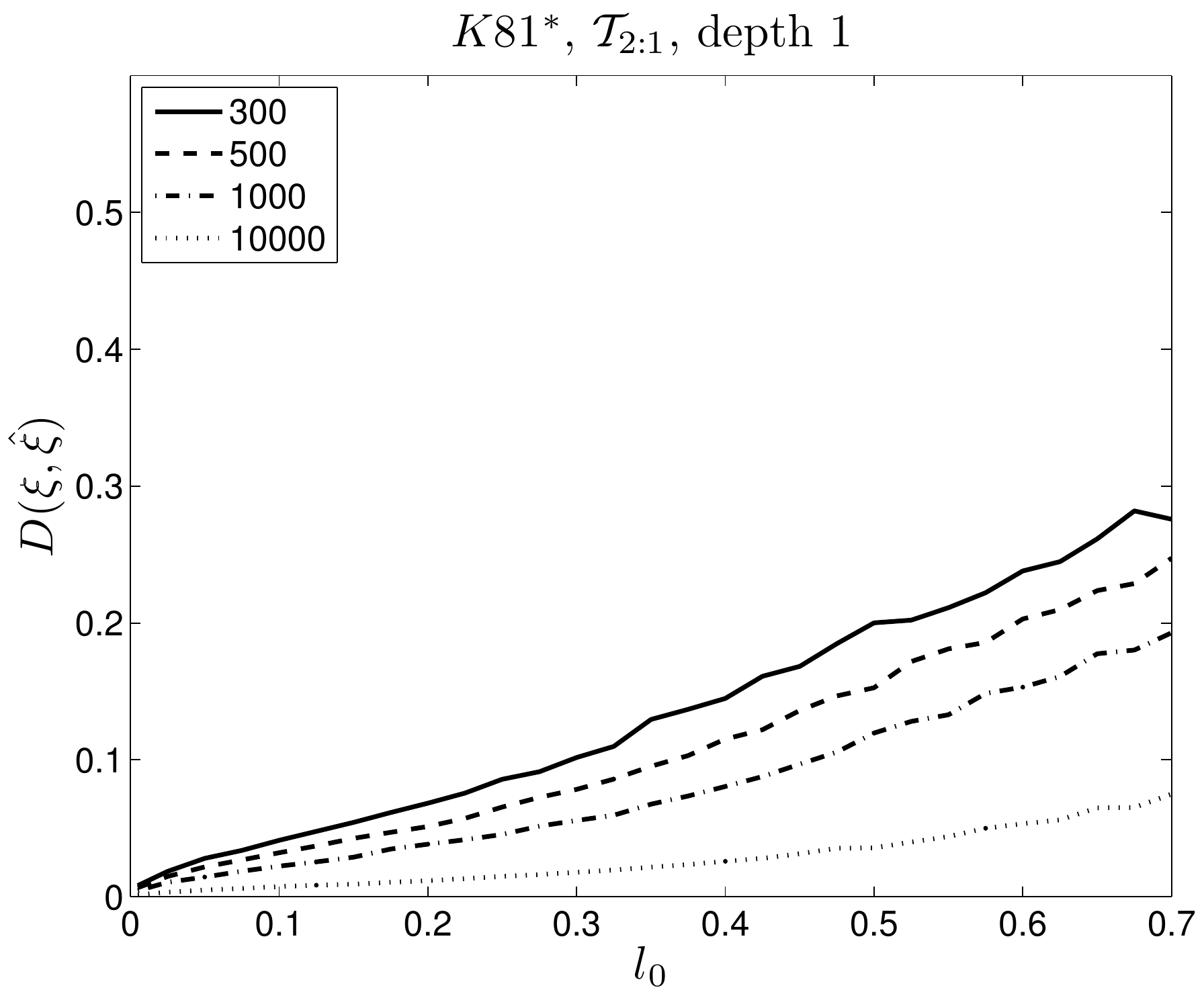}
\includegraphics[scale=0.18]{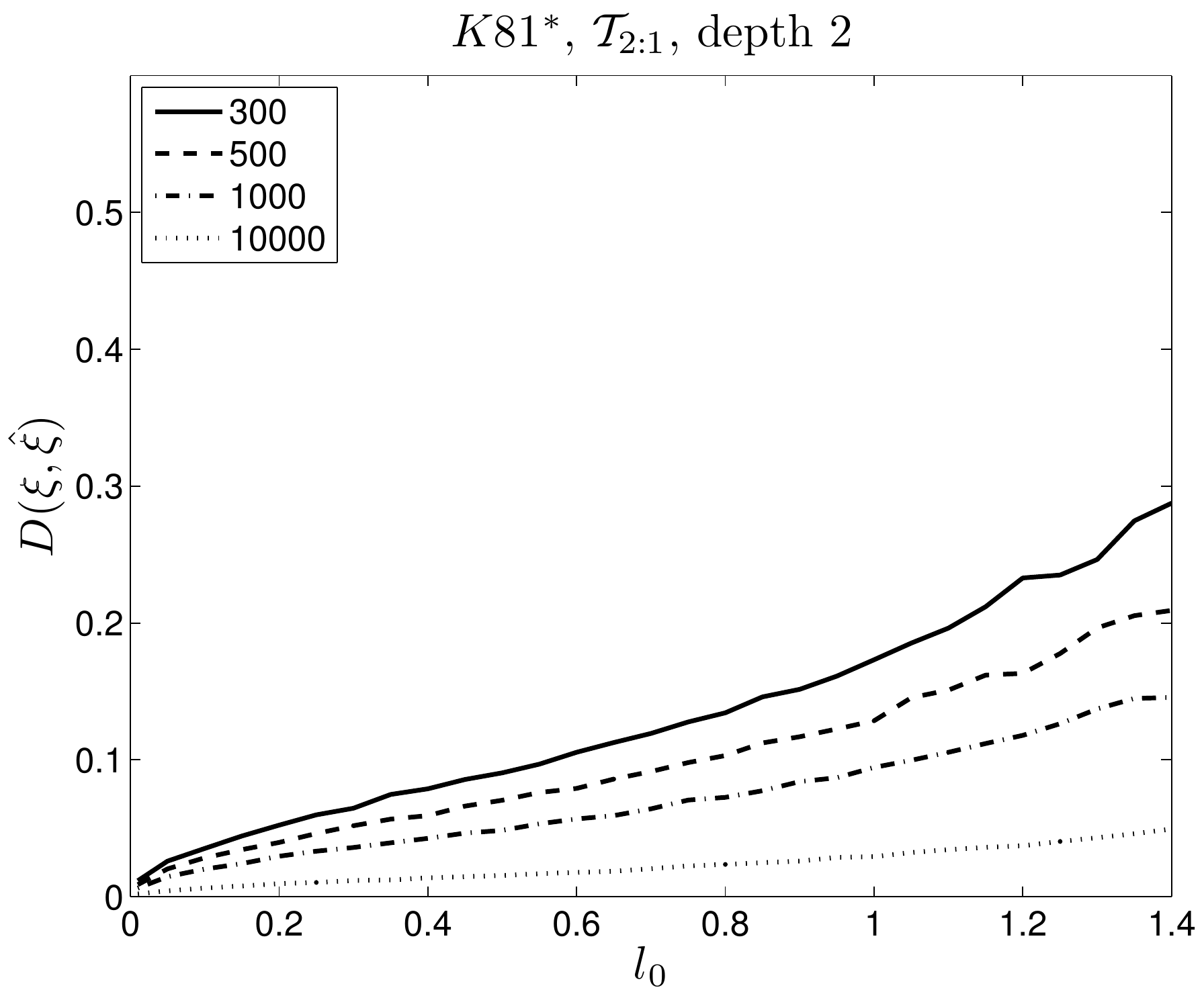}\vfill
\includegraphics[scale=0.18]{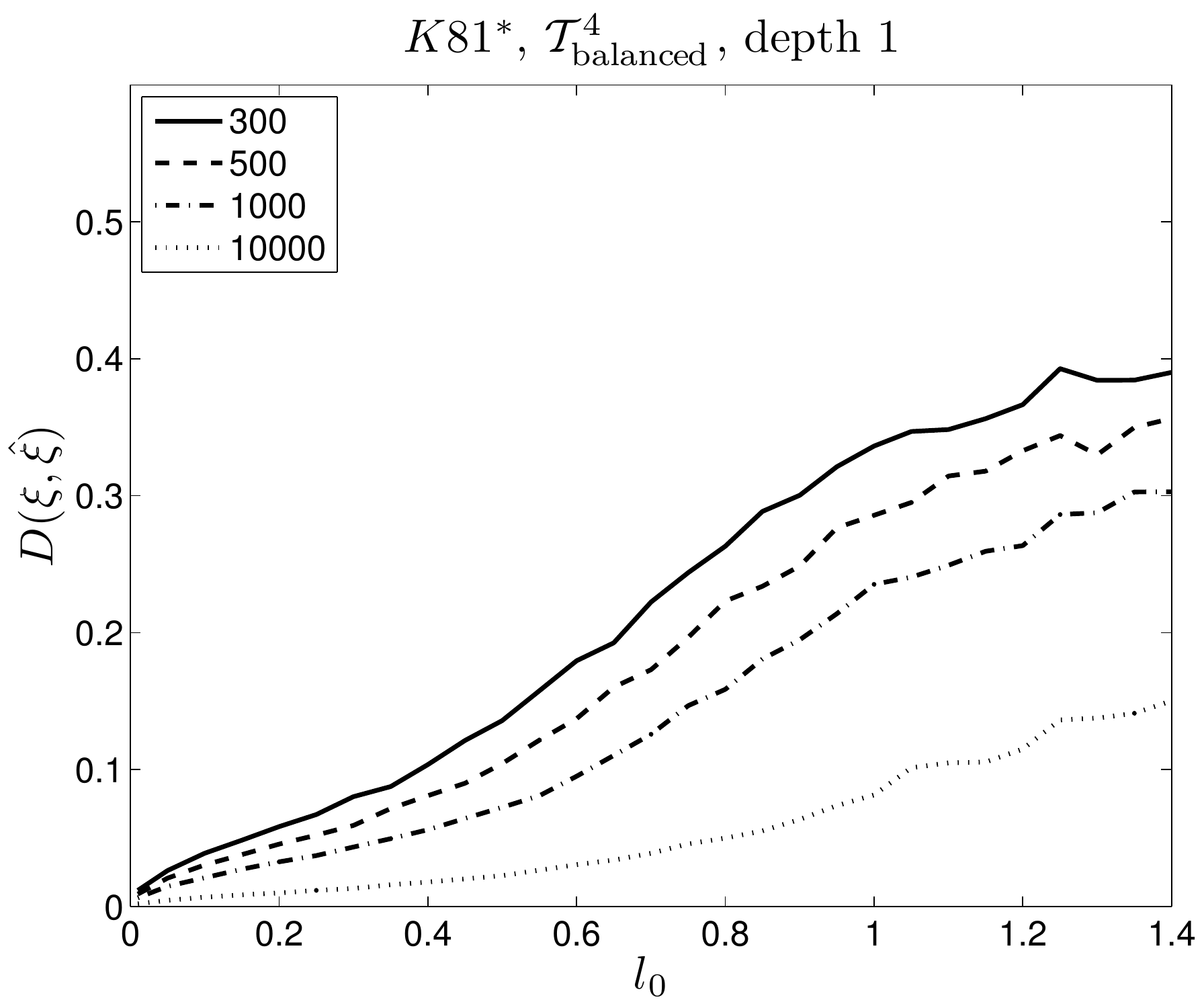}
\includegraphics[scale=0.18]{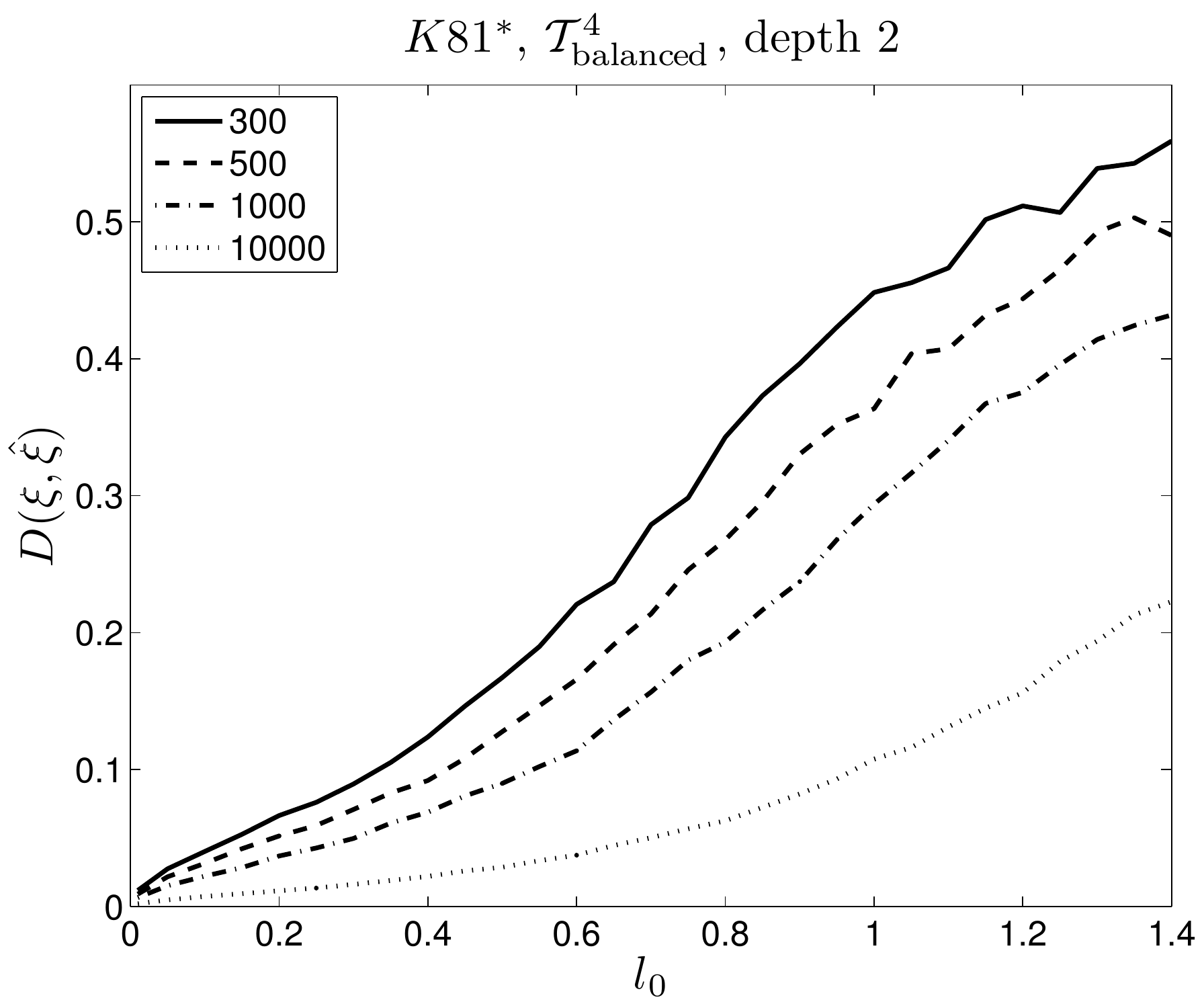}
\end{minipage}\label{fig:k81_dist}}
\caption{\small Divergence $D(\xi,\hat{\xi})$ between the
parameters, $\xi$, and their estimates, $\hat{\xi}$, calculated by
$\empar$. Horizontal axis: original length
of the inner branch.}\label{fig:L1ind}
\end{figure}

\begin{figure}[h!]
 \begin{minipage}[b]{0.7\linewidth}
\subfigure{
\includegraphics[scale=0.35]{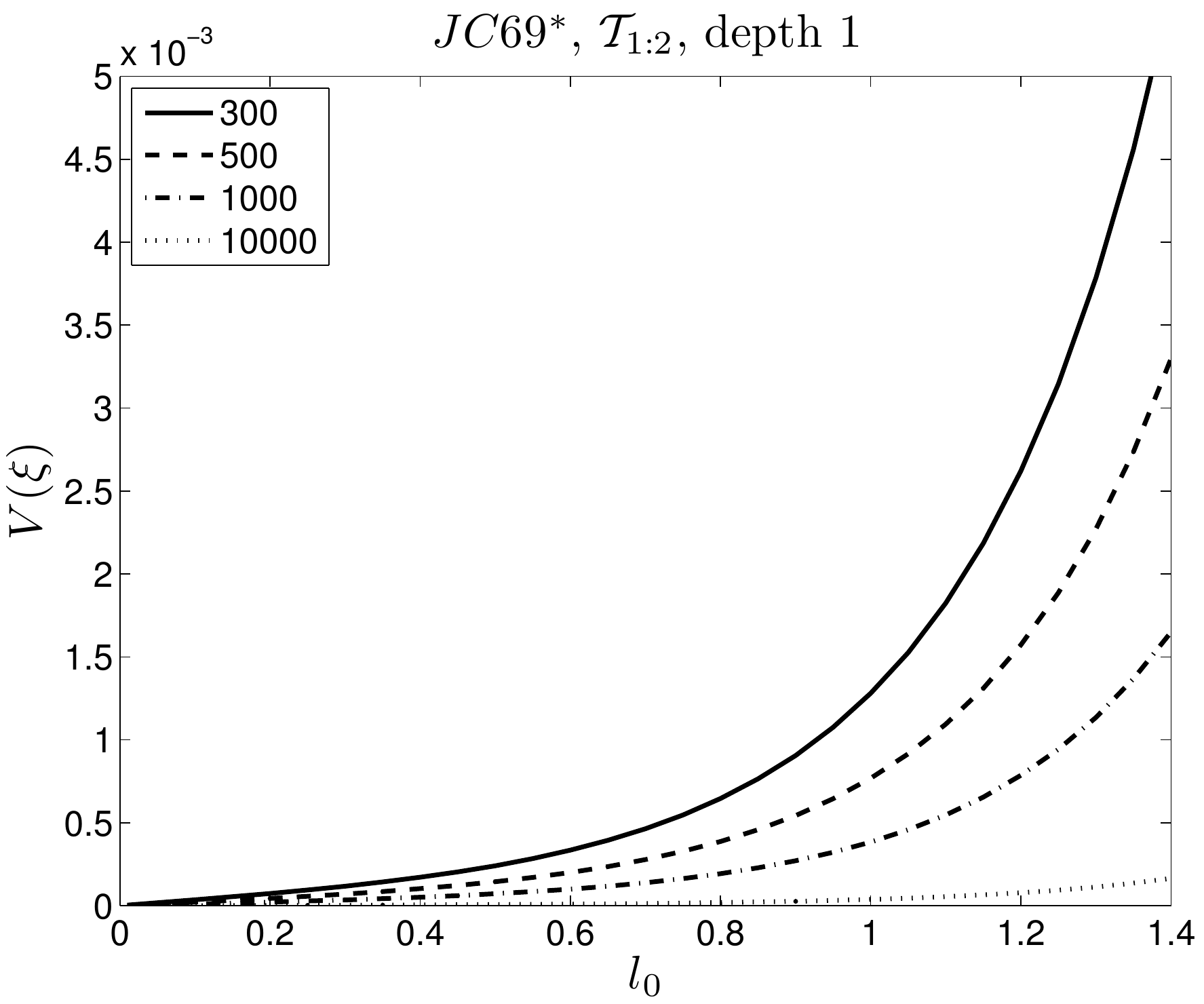}
\includegraphics[scale=0.35]{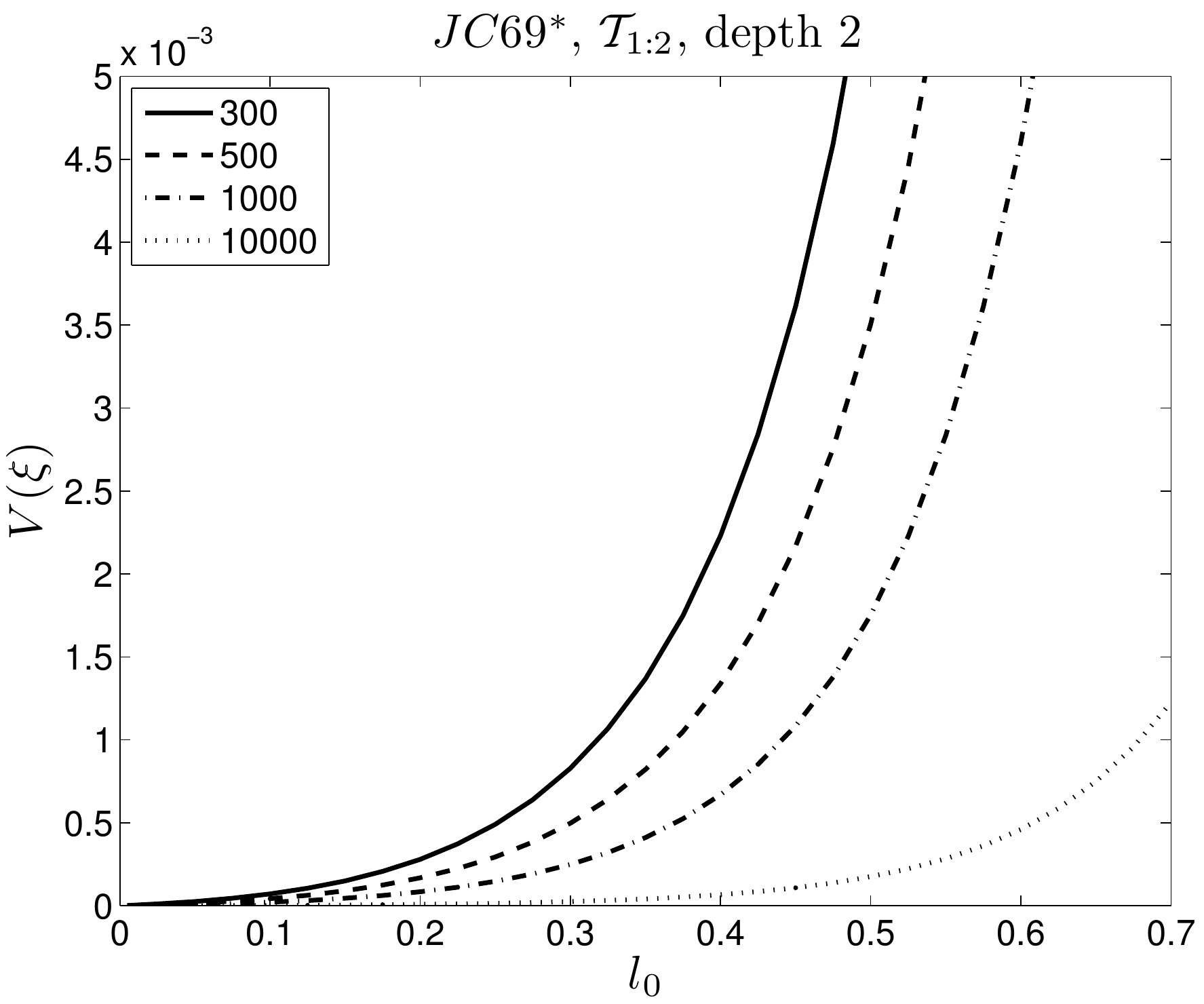}}
 \subfigure{\includegraphics[scale=0.35]{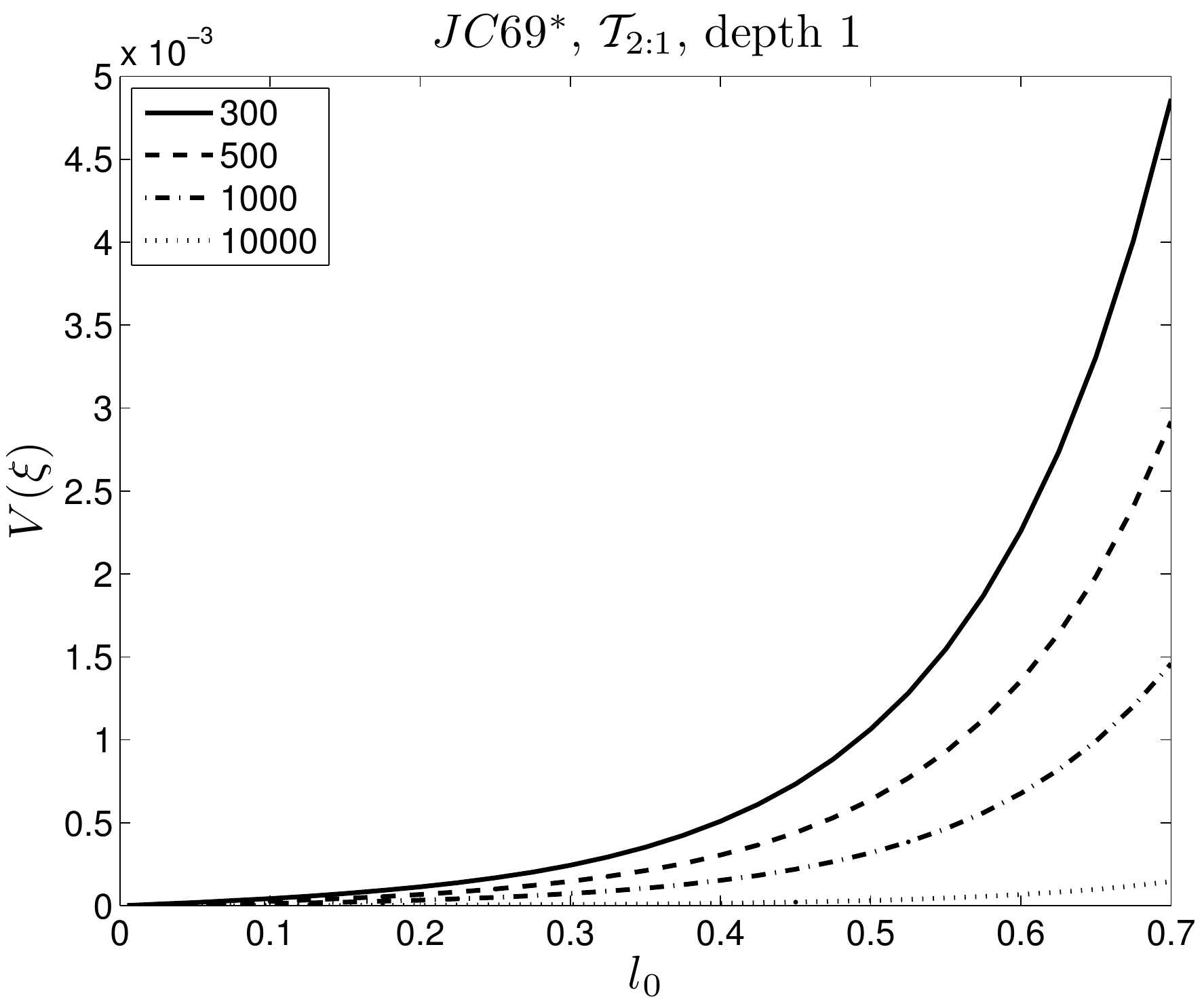}
\includegraphics[scale=0.35]{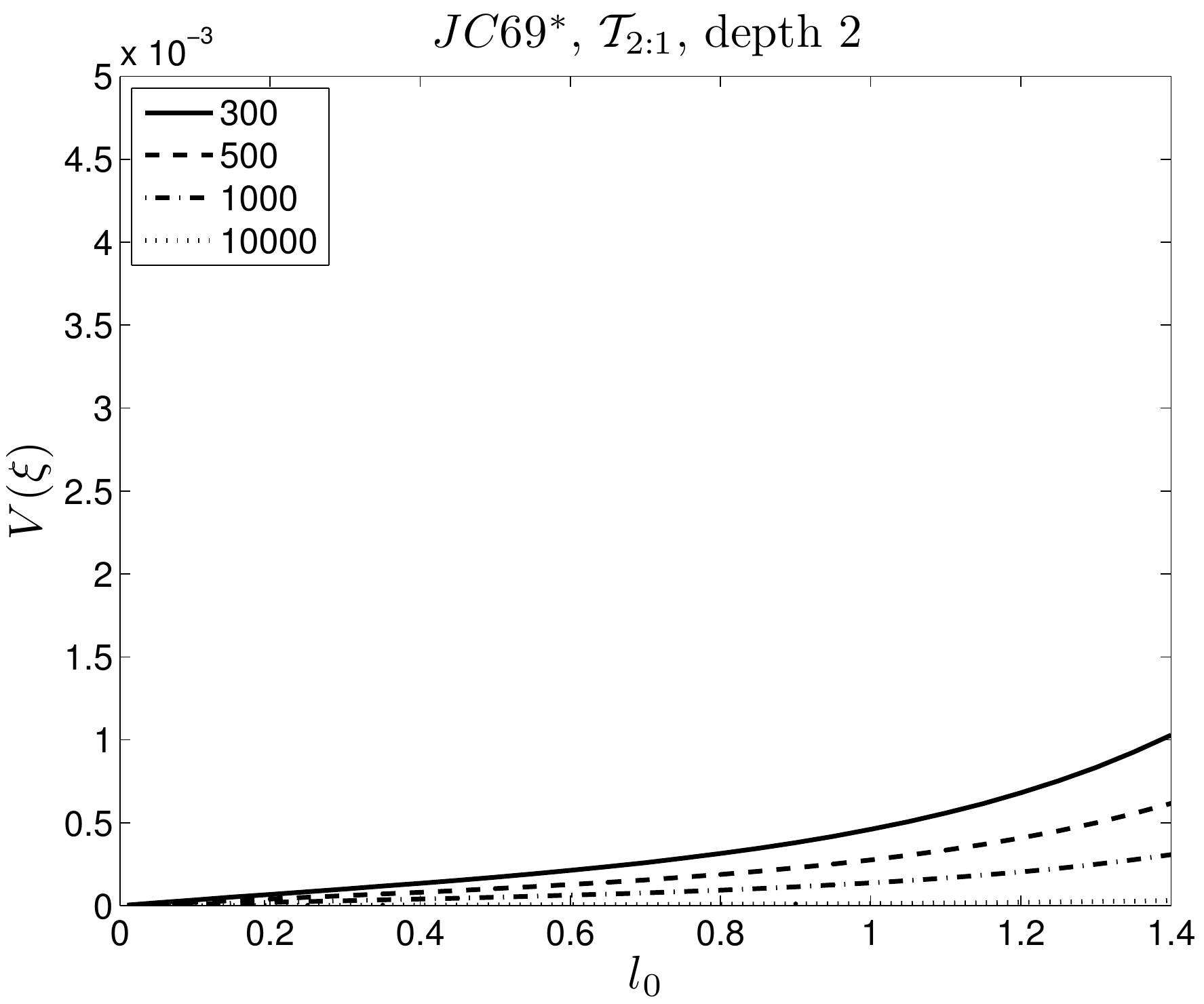}}
\subfigure{
\includegraphics[scale=0.35]{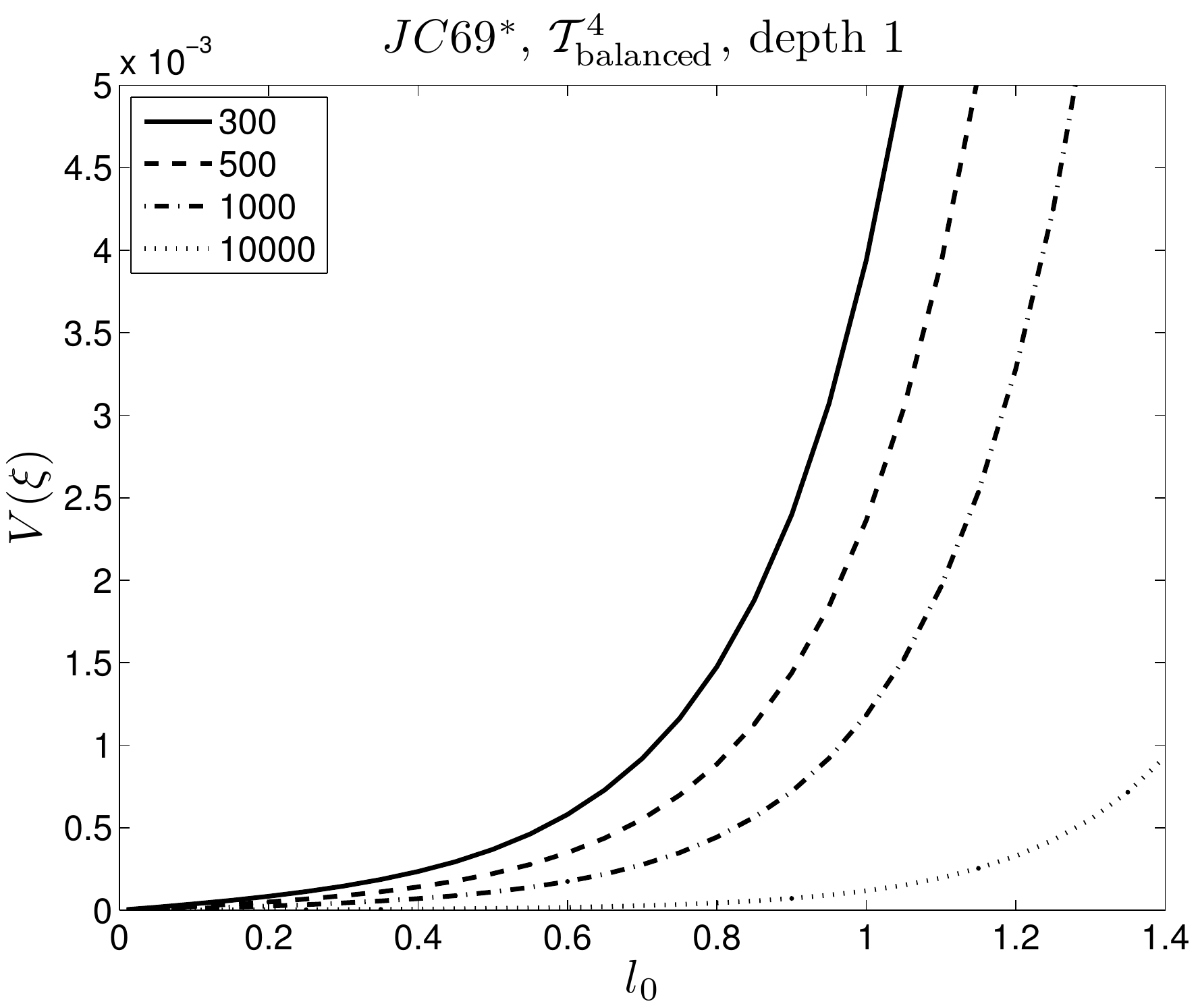}
\includegraphics[scale=0.35]{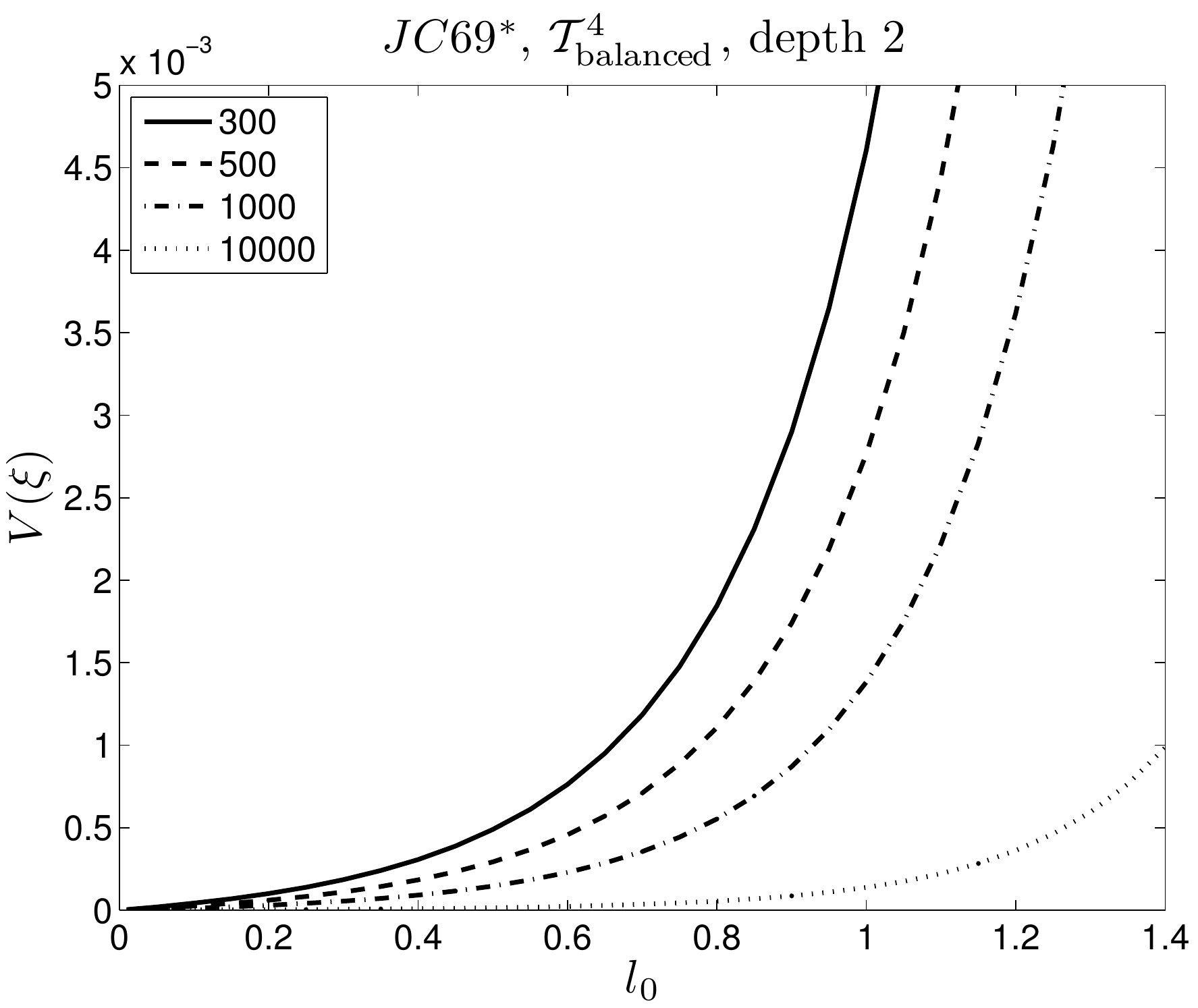}}
\end{minipage}
\caption{\small Distribution of variances of the estimated
parameters for different alignment lengths
 and different lengths of the depth 1 (\emph{left}) and depth 2 (\emph{right}) branches under the $\jc$ model: $\stwo$ (\emph{top}),  $\sth$  (\emph{middle}),  $\sone^4$ (\emph{bottom}).}\label{fig:jc_var}
\end{figure}

\begin{figure}[h!]
 \begin{minipage}[b]{0.7\linewidth}
\subfigure{
\includegraphics[scale=0.25]{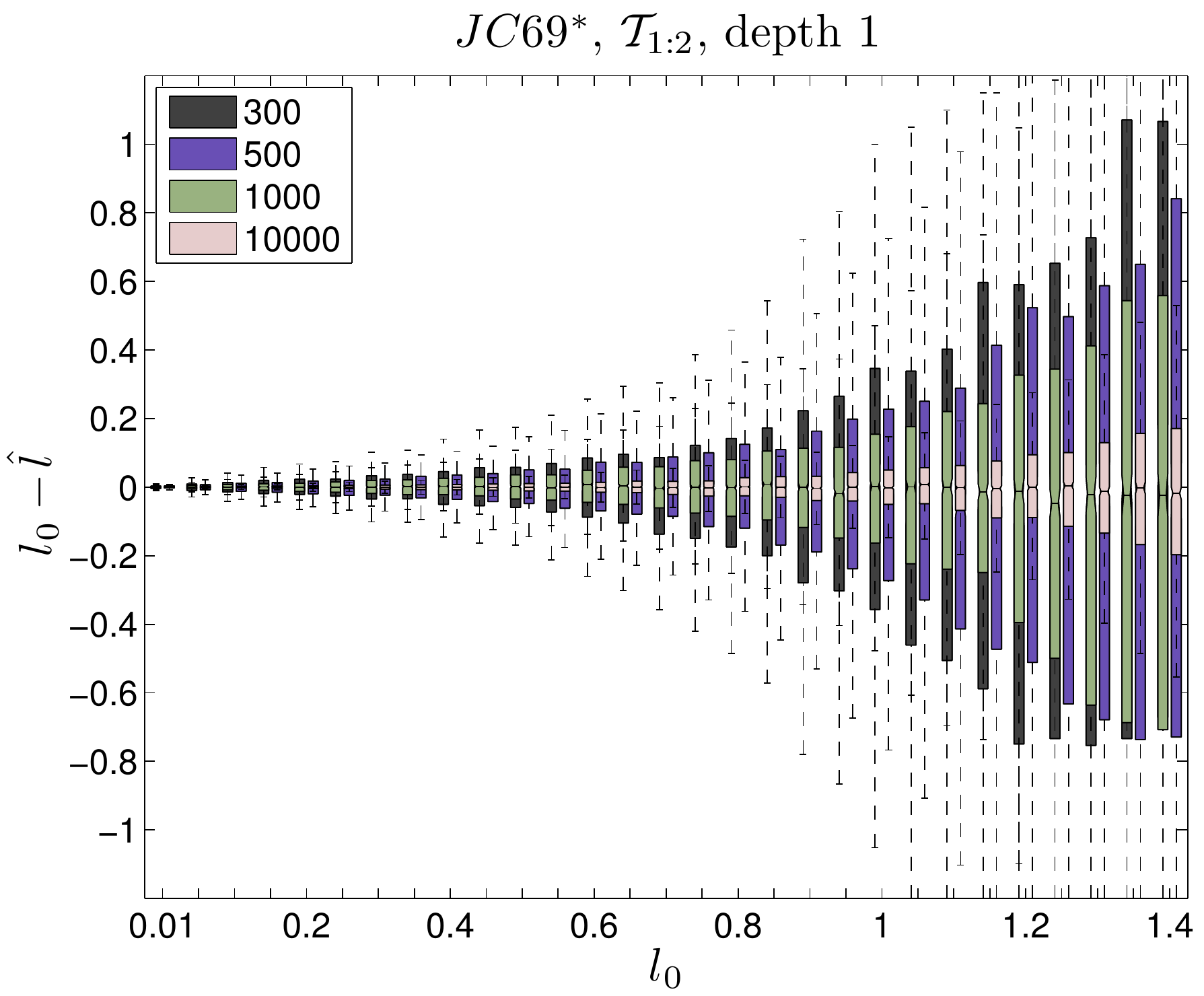}\hspace{1mm}
\includegraphics[scale=0.25]{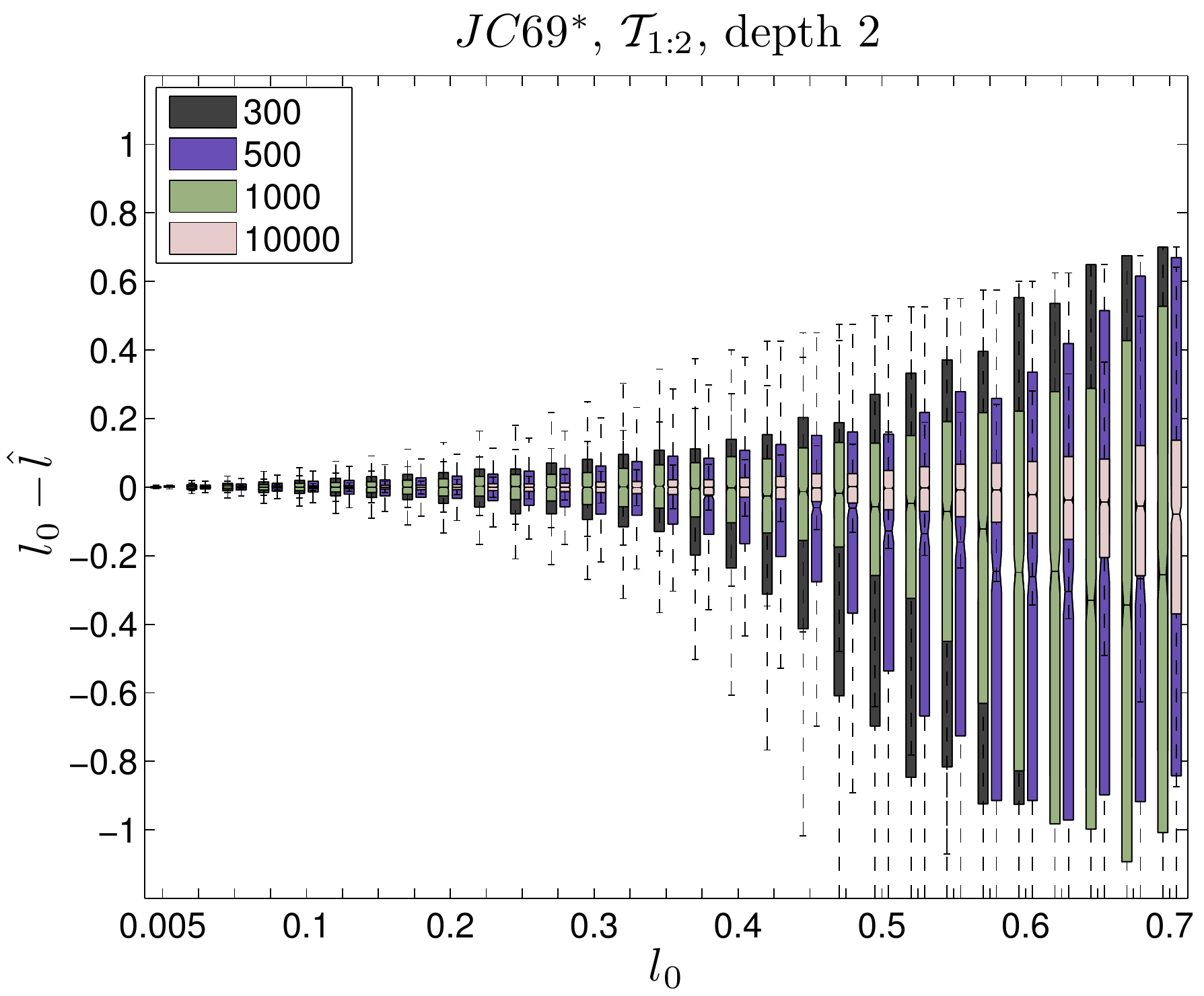}
\includegraphics[scale=0.25]{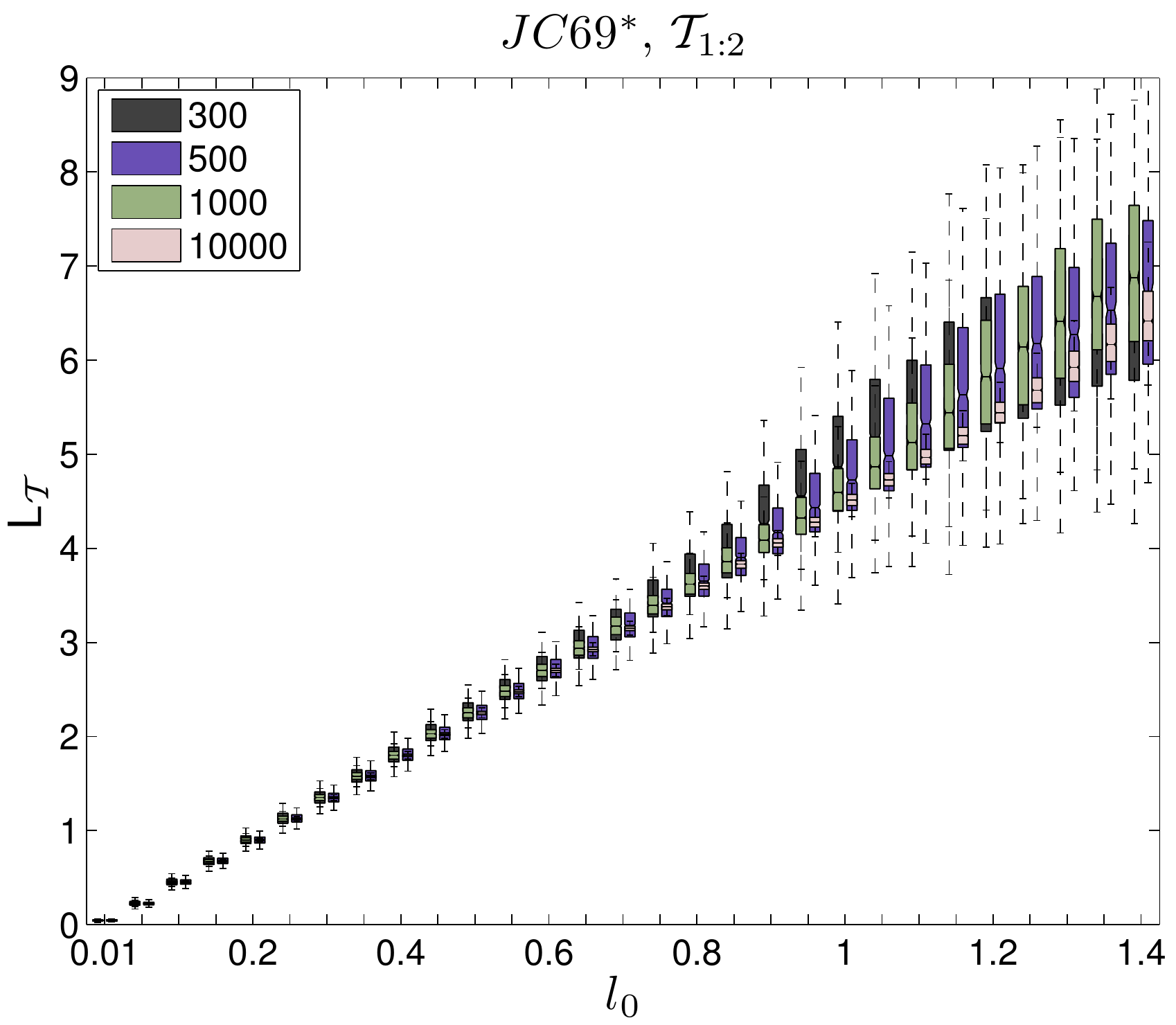}}
\subfigure{
\includegraphics[scale=0.25]{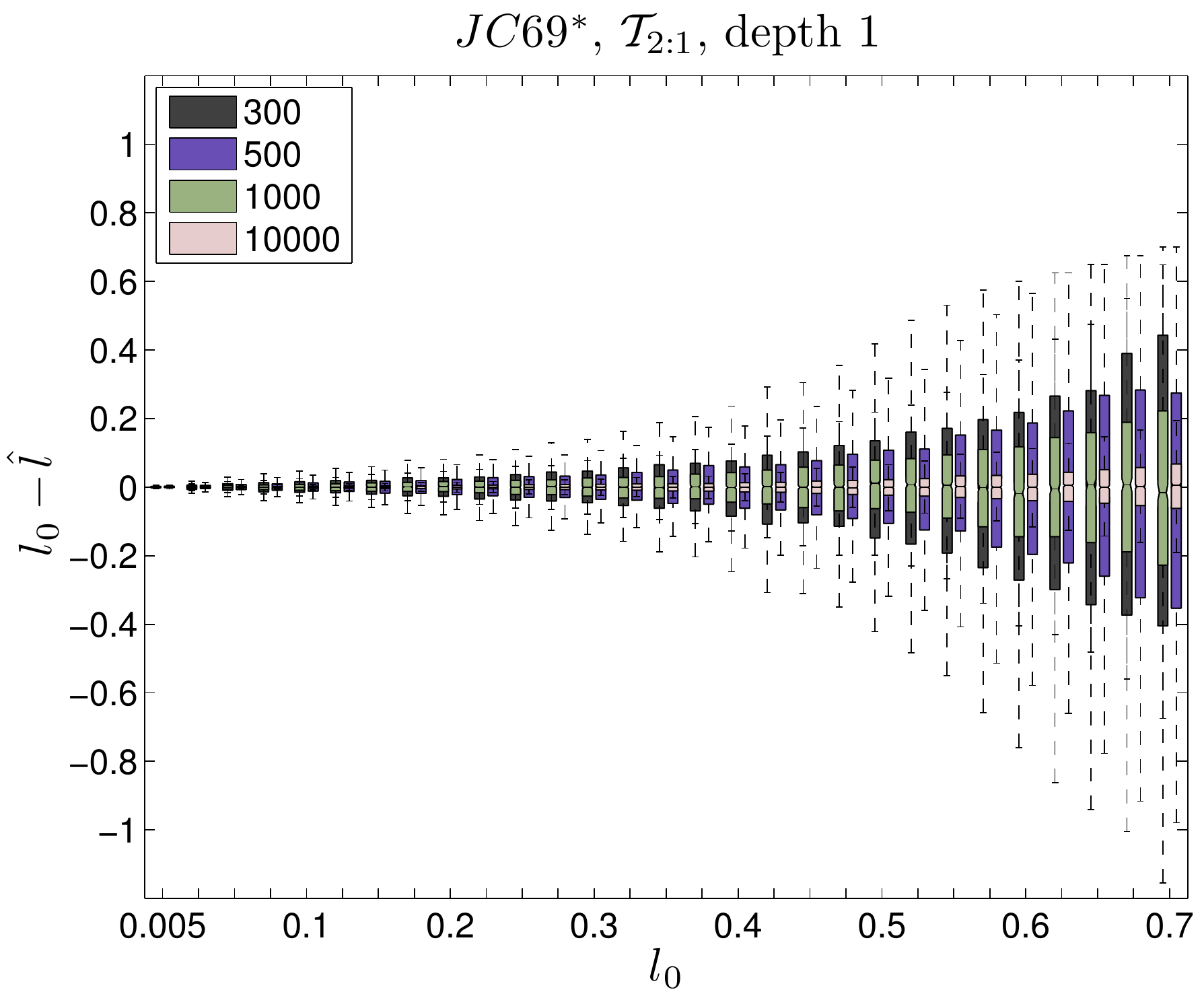}\hspace{1mm}
\includegraphics[scale=0.25]{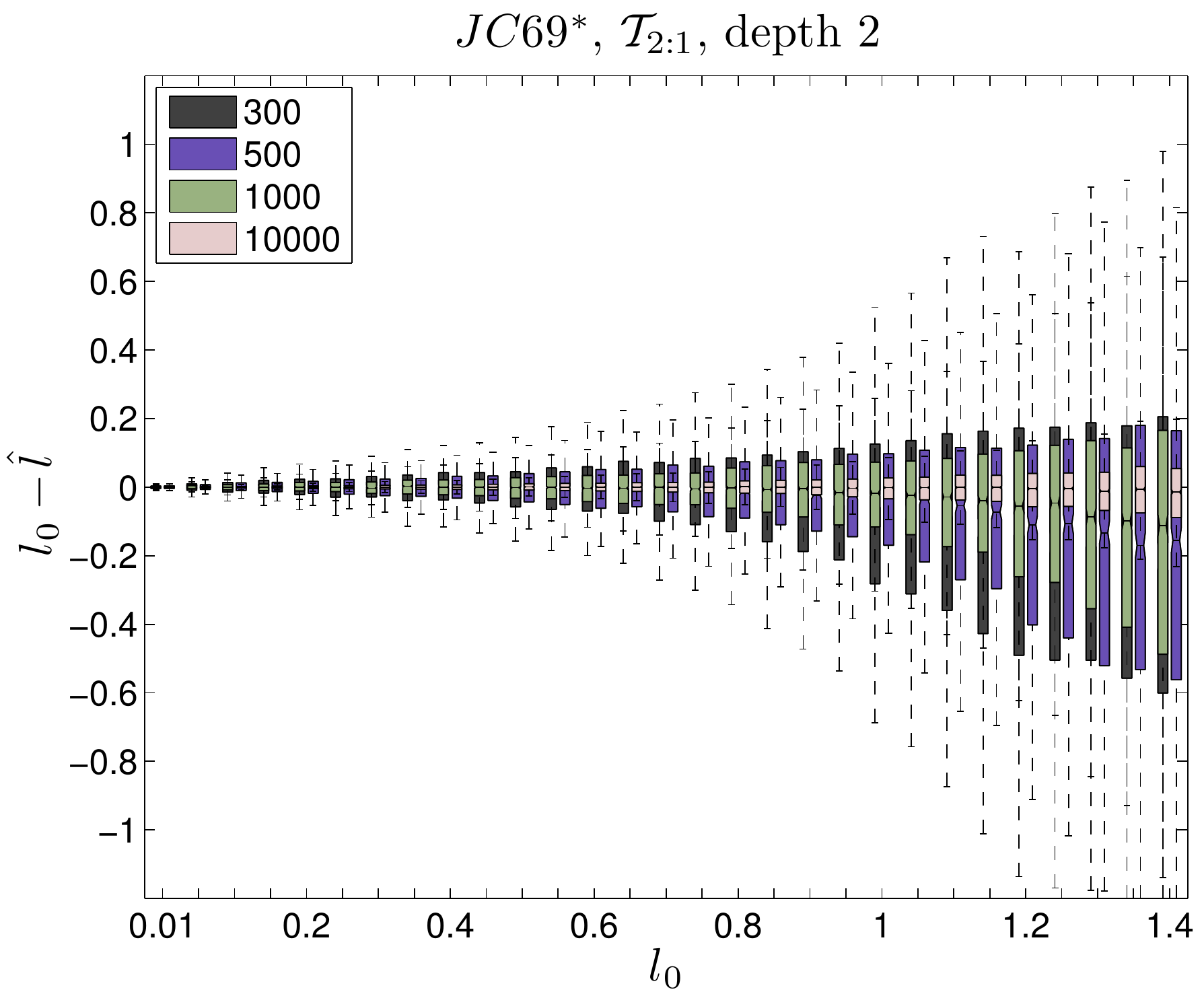}
\includegraphics[scale=0.25]{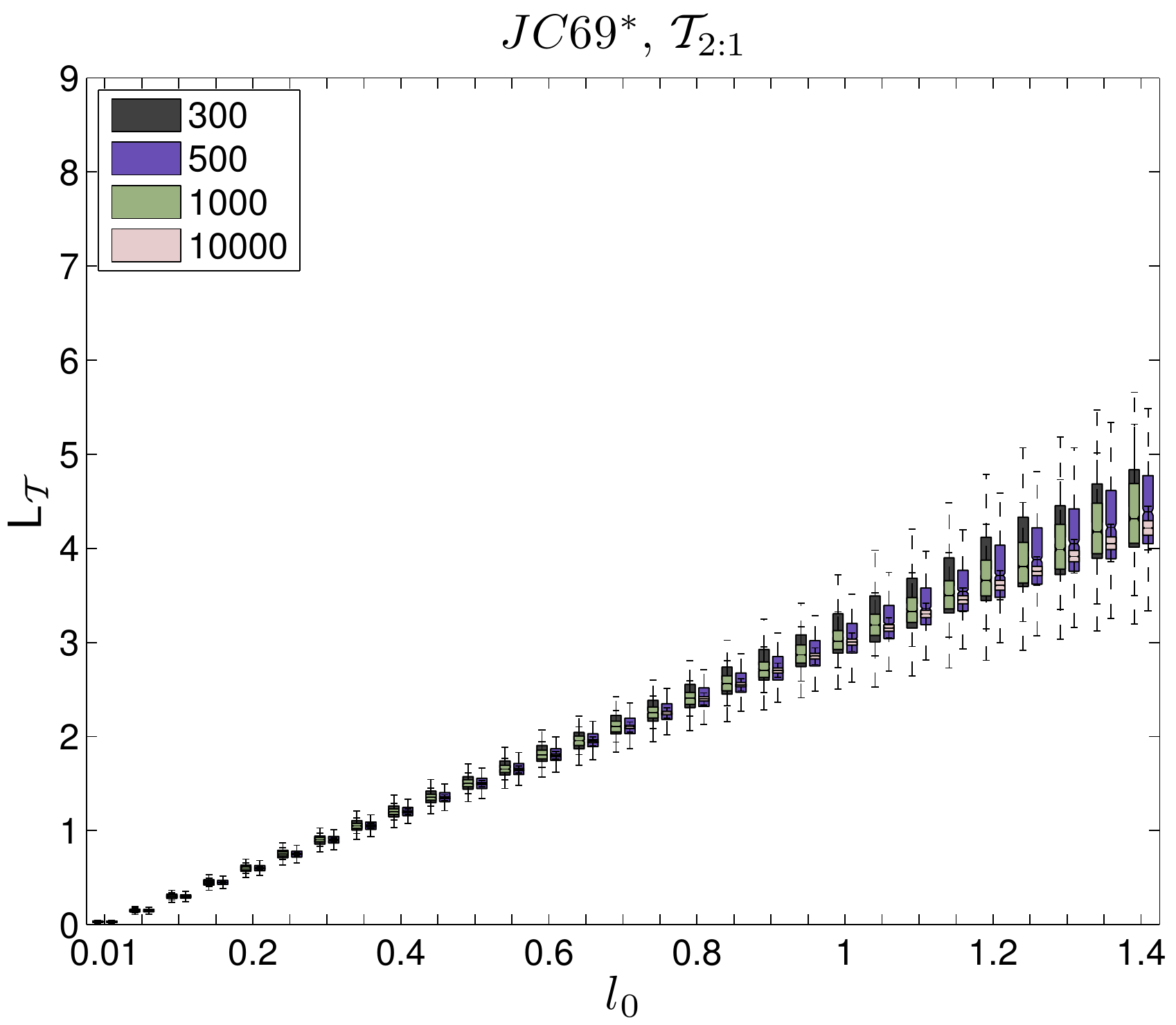}}
\subfigure{
\includegraphics[scale=0.25]{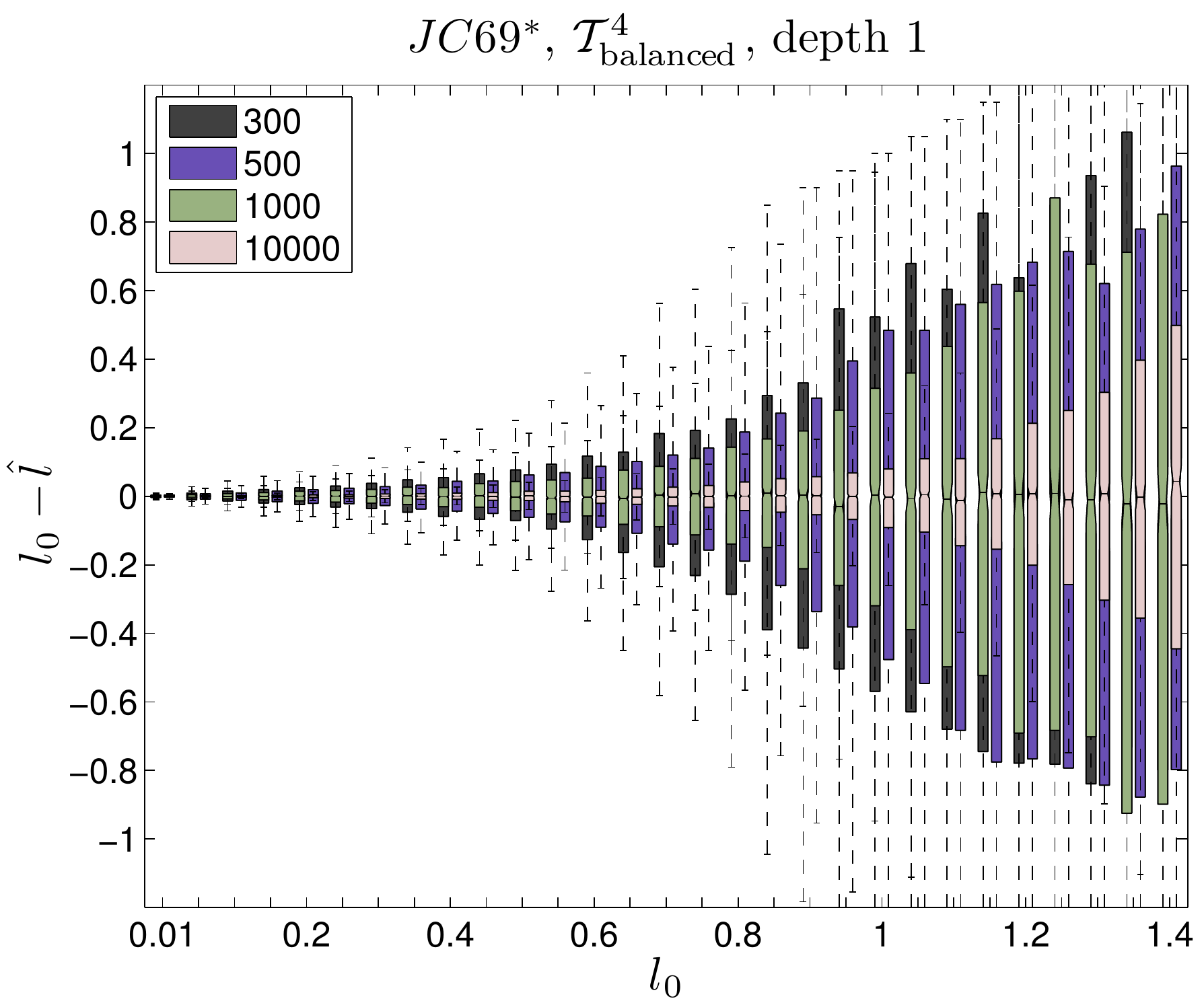}\hspace{1mm}
\includegraphics[scale=0.25]{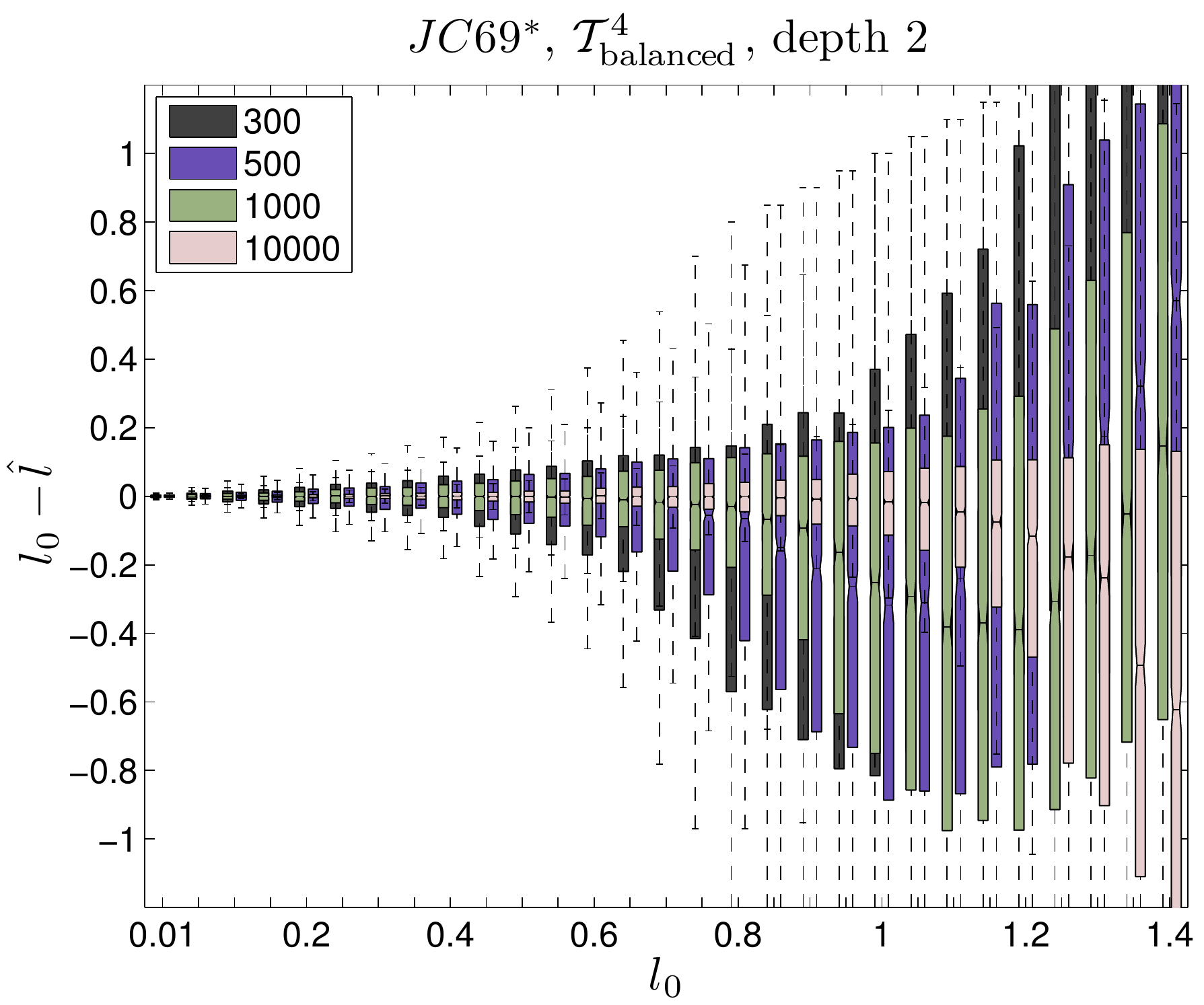}
\includegraphics[scale=0.25]{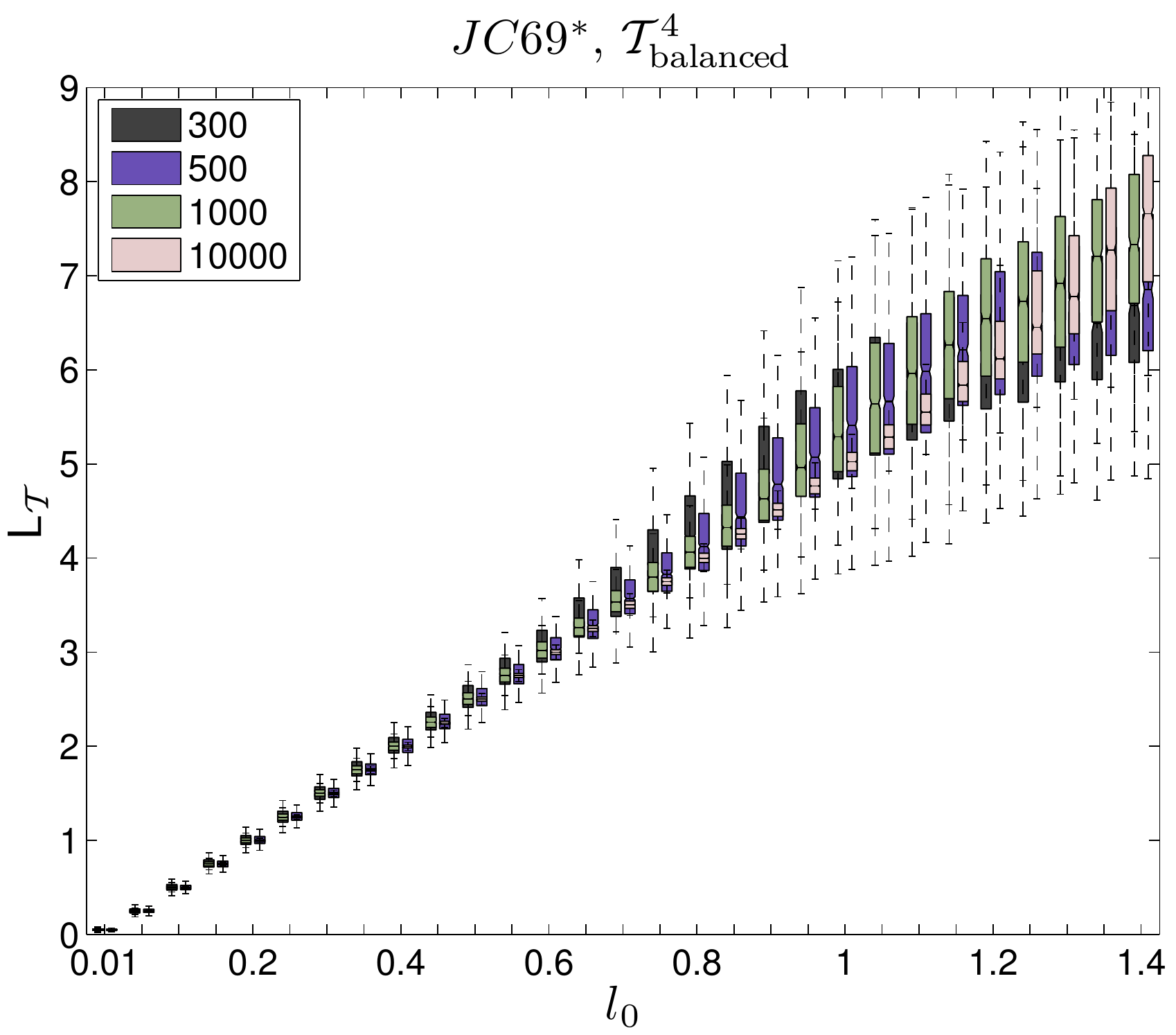}}
\end{minipage}
\caption{\small Error in the branch length estimation measured as the
  difference between the initial and the estimated branch
  lengths, $l_0-\hat{l}$, in the
$1,000$ simulated data sets along the $\sone^4,
  \stwo, \sth$ trees under the $\jc$ model (\emph{left and middle
    columns}). \emph{Rightmost column} displays the distribution of
  the estimated length of the tree, where $l_0$ labeling the
  horizontal axis corresponds to the length
  of the internal branch in $\tree$.}\label{fig:jc_br}
\end{figure}

\begin{figure}[h!]
 \begin{minipage}[b]{0.7\linewidth}
\subfigure{
\includegraphics[scale=0.25]{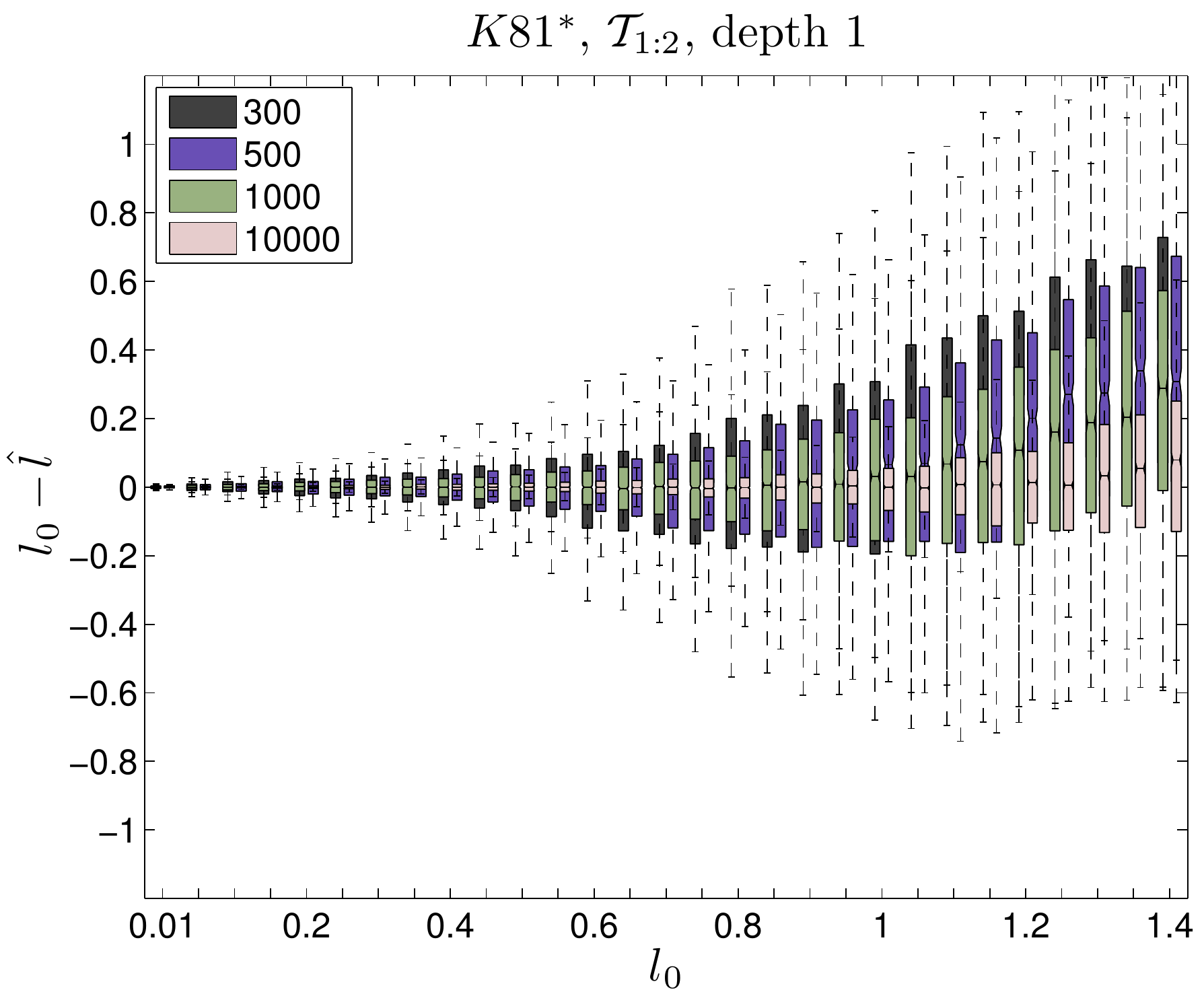}\hspace{1mm}
\includegraphics[scale=0.25]{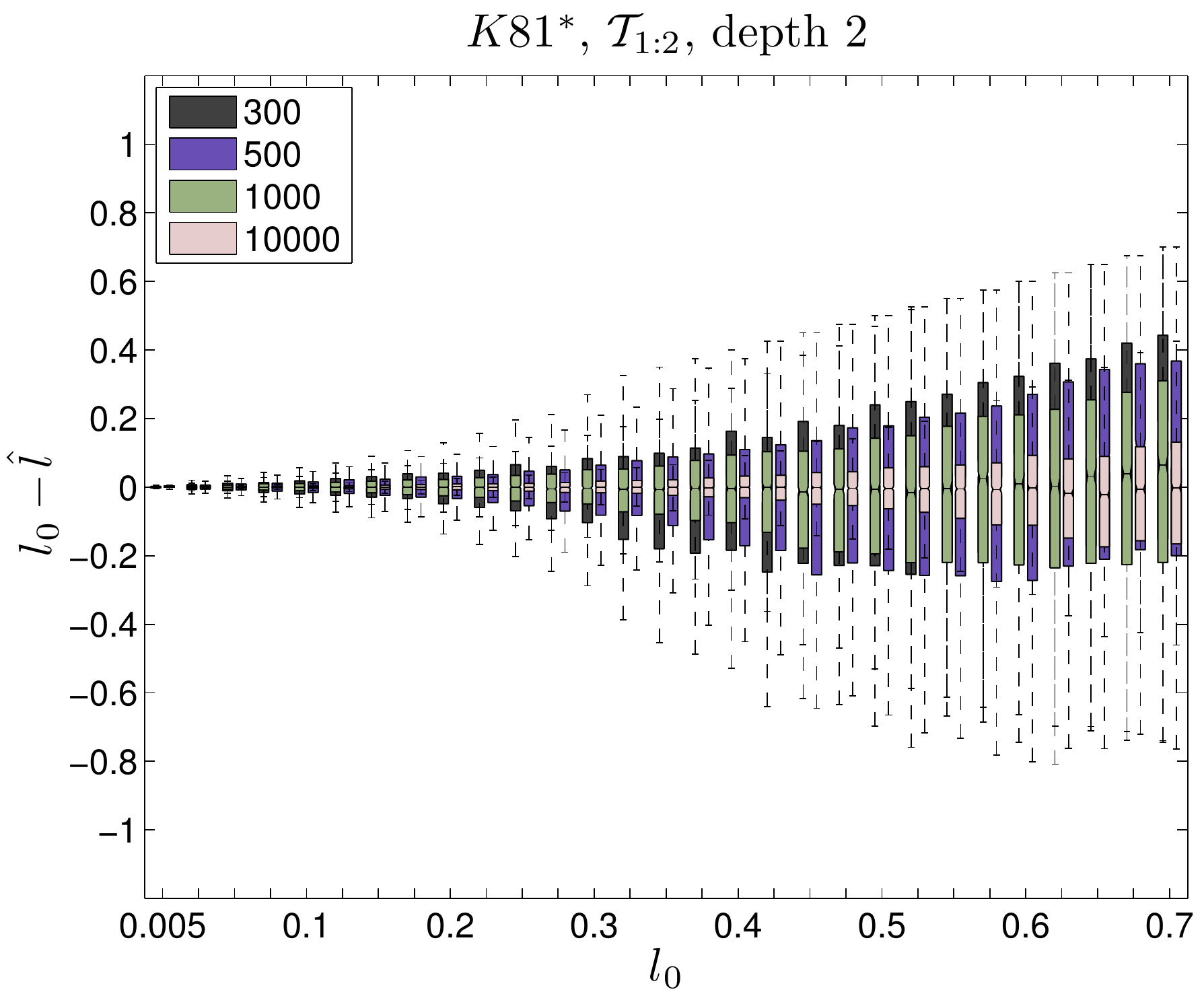}
\includegraphics[scale=0.25]{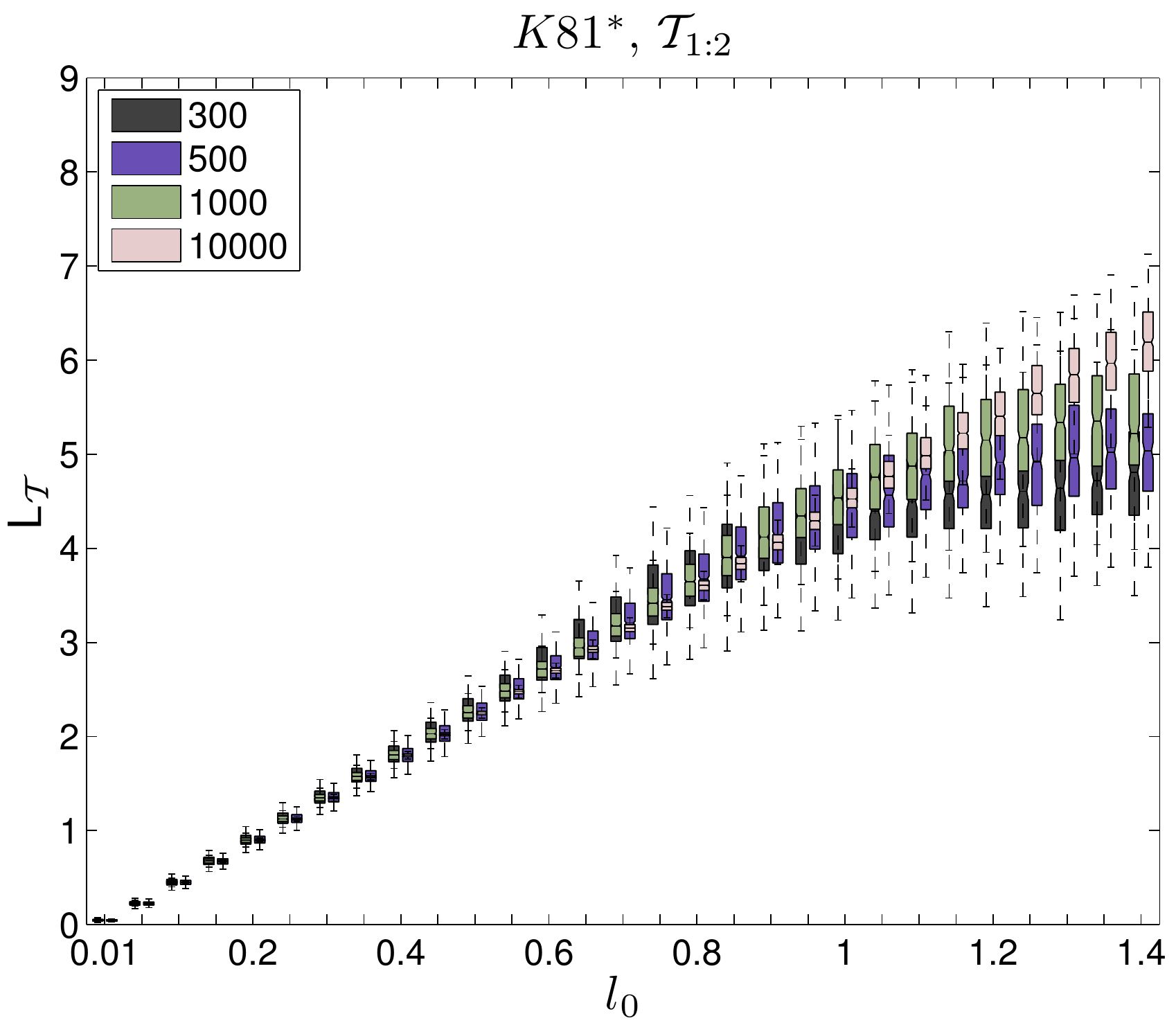}}
\subfigure{
\includegraphics[scale=0.25]{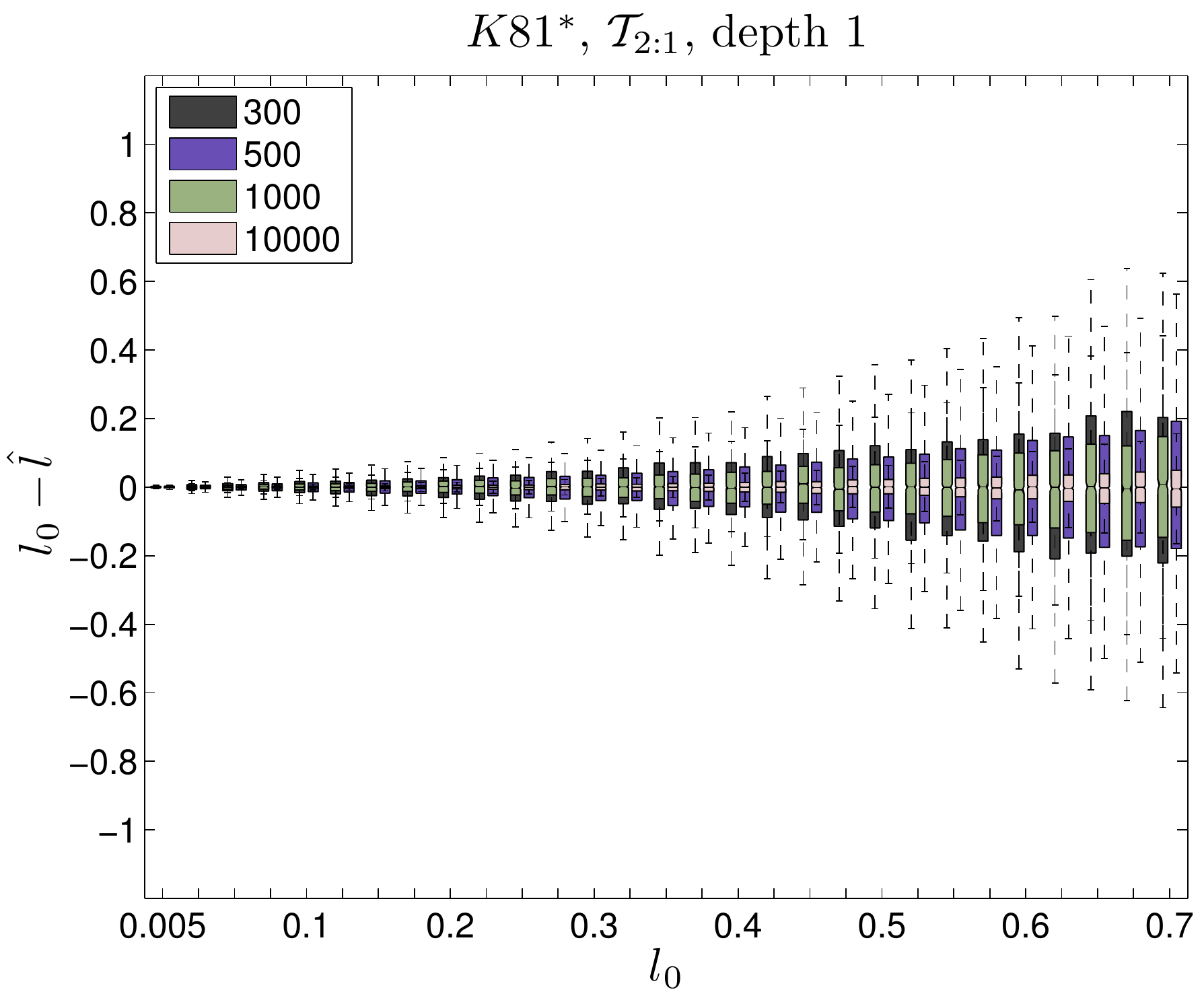}
\includegraphics[scale=0.25]{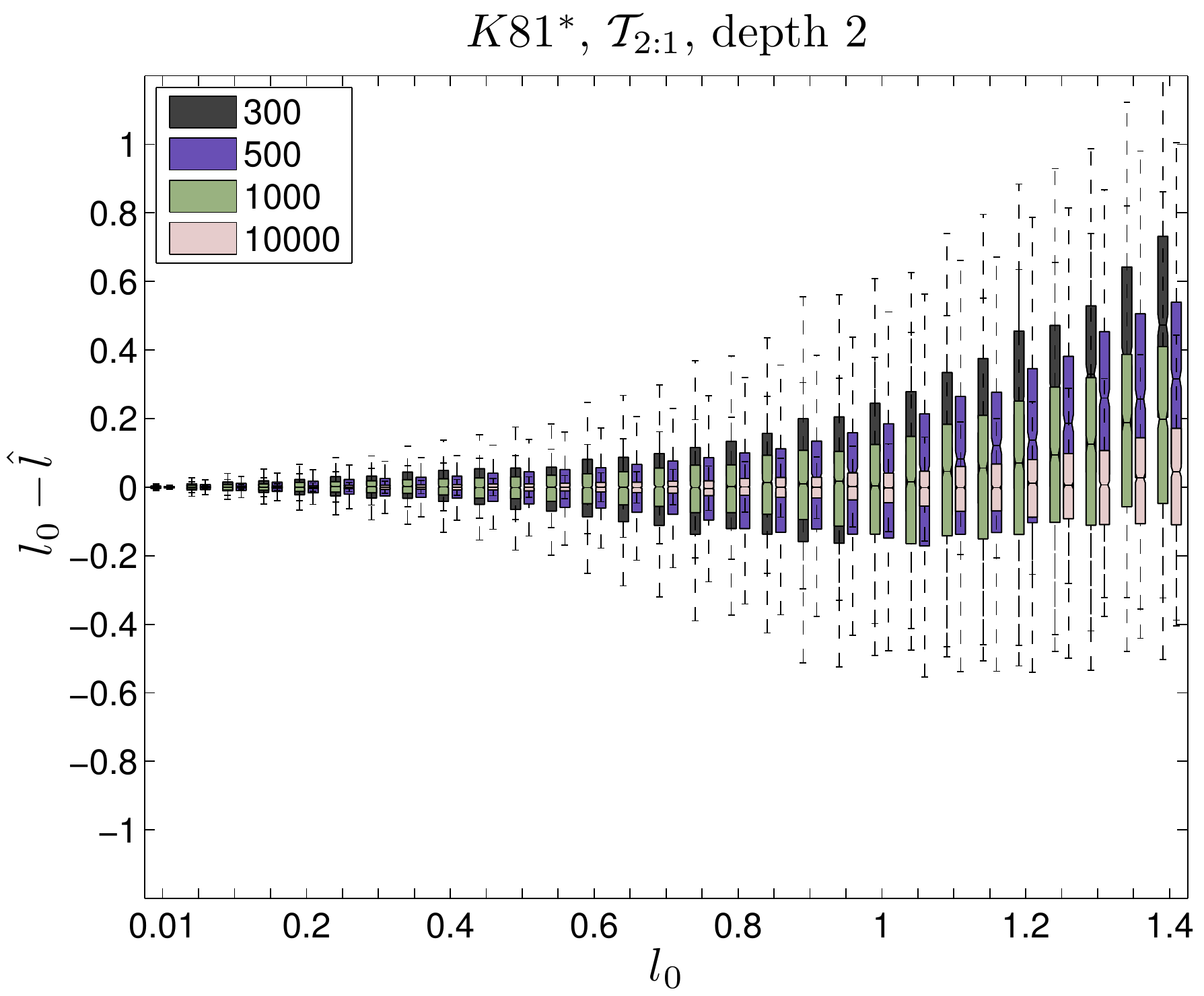}
\includegraphics[scale=0.25]{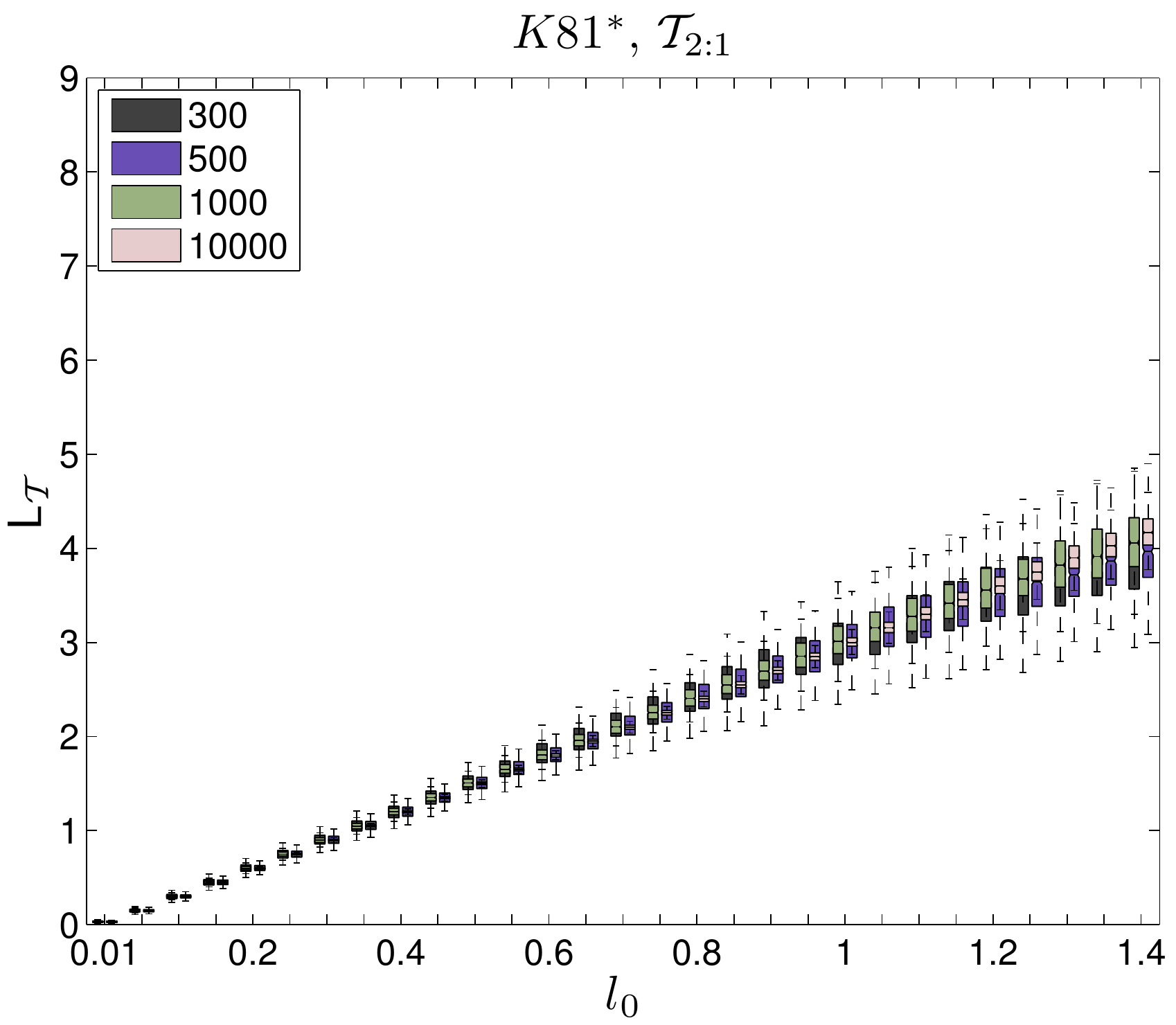}}
\subfigure{
\includegraphics[scale=0.25]{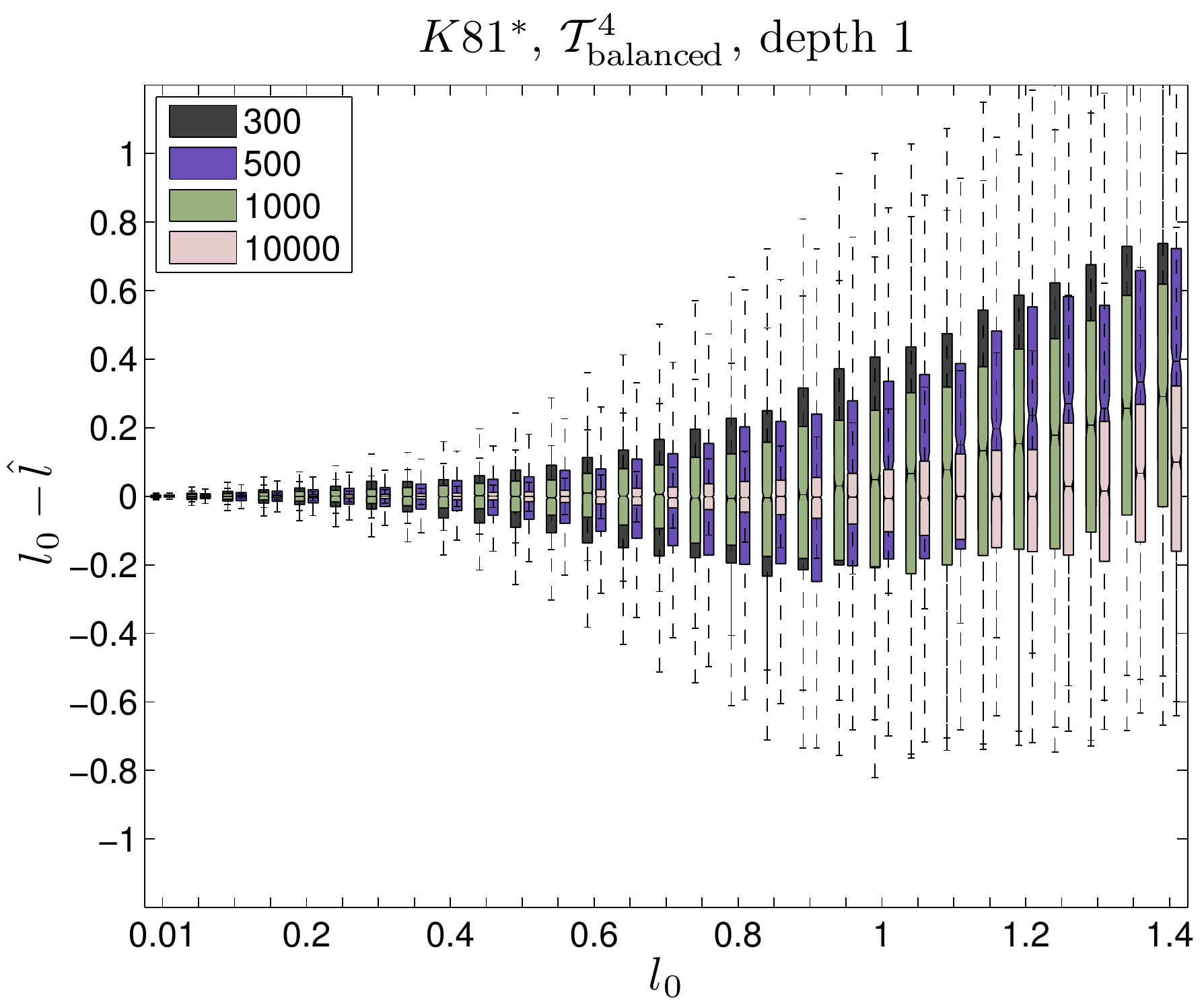}\hspace{1mm}
\includegraphics[scale=0.25]{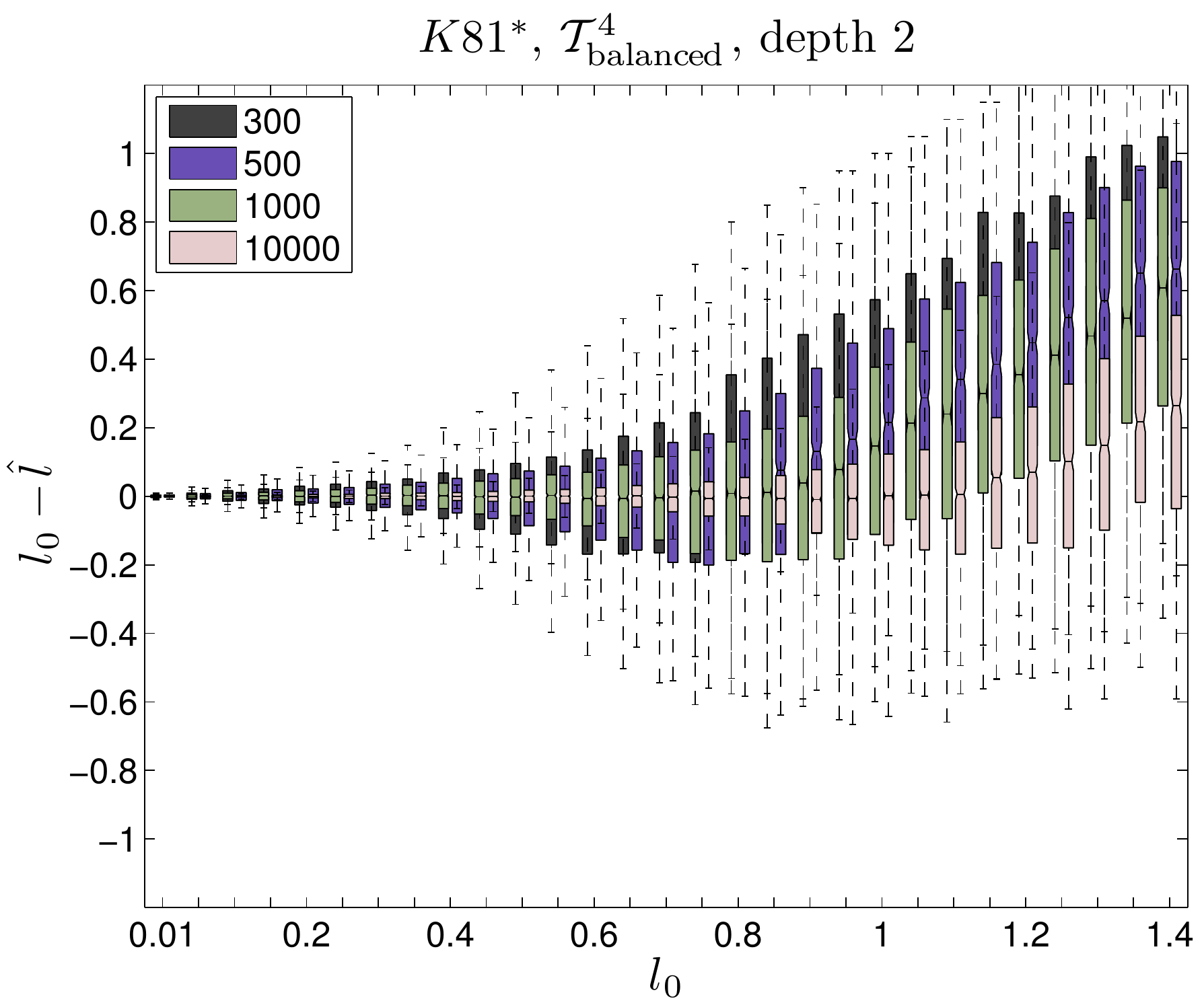}
\includegraphics[scale=0.25]{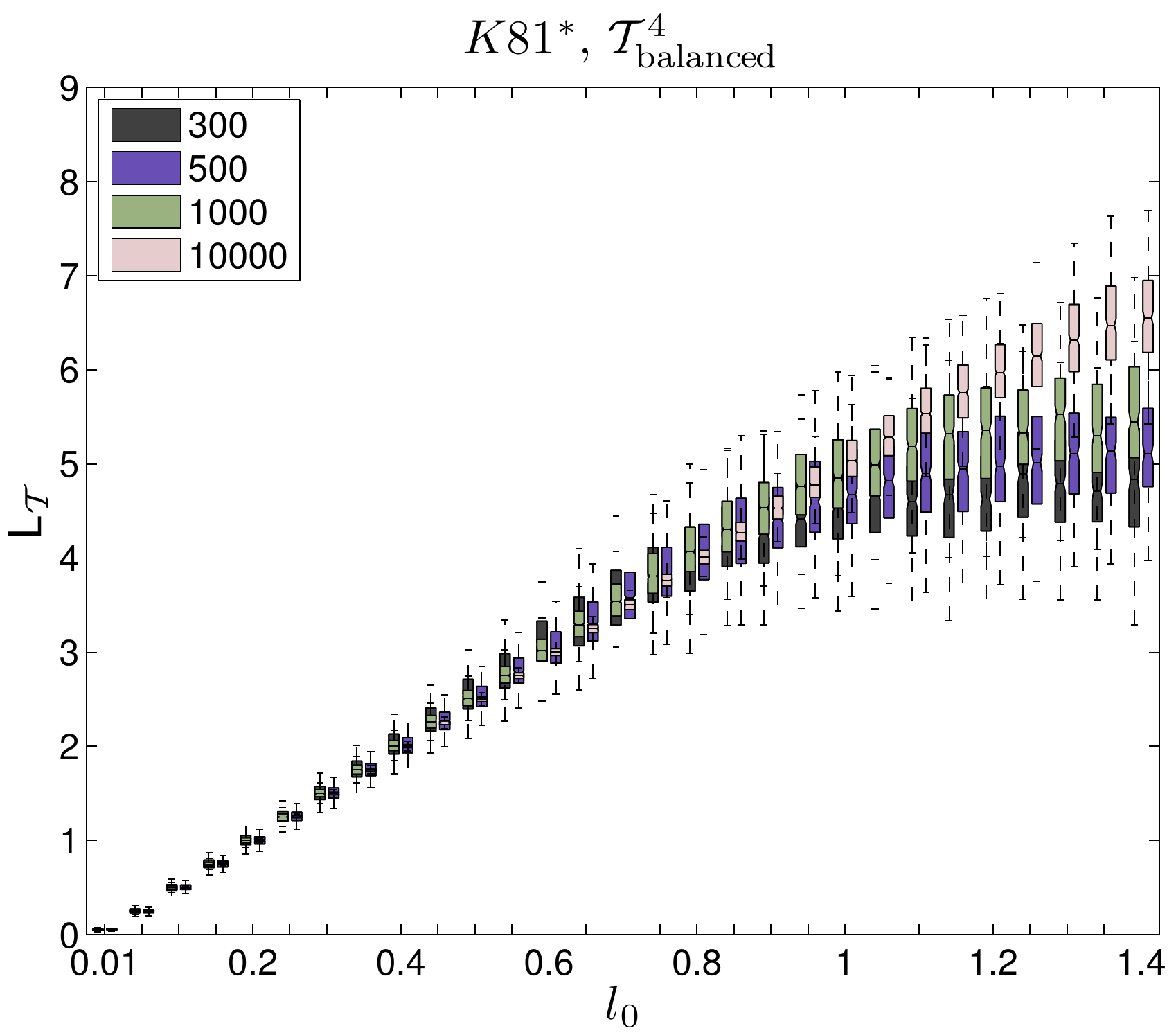}}
\end{minipage}
\caption{Error in the branch length estimation under the $\kth$ model
  (see Fig. 4 
 for details).}\label{fig:k81_br}
\end{figure}

\begin{figure}[h!]
 \begin{minipage}[b]{0.7\linewidth}
\hspace{35mm}
\includegraphics[scale=0.38]{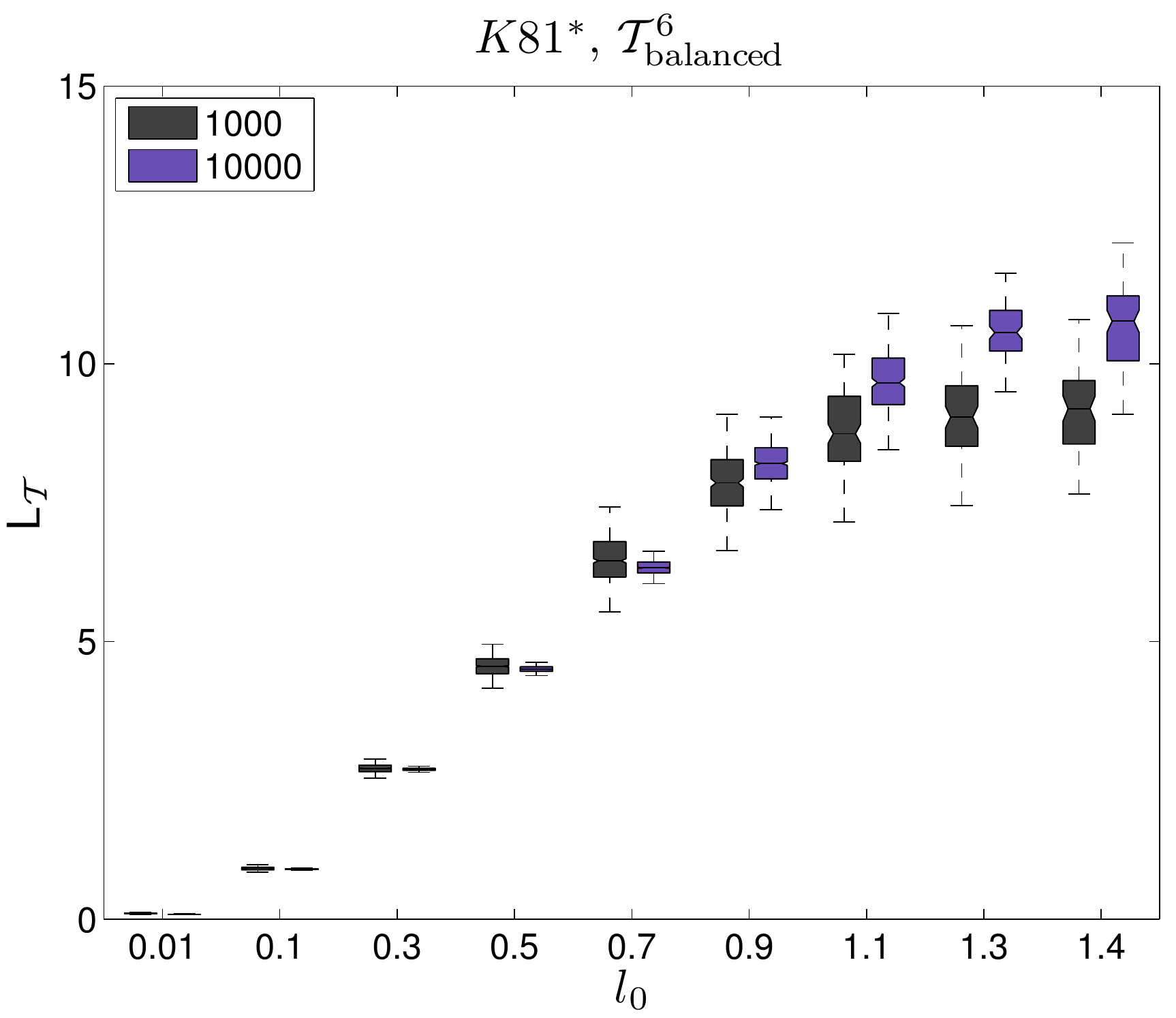}
\end{minipage}
\caption{\small Estimated tree length as a function of the
initial length of a branch of $\sone^6$  ($\len=9l_0$) in $1,000$ data sets generated under the $\kth$ model.}\label{fig:sixtaxa_len}
\end{figure}

\thispagestyle{empty}
\begin{figure}[h!]
 \begin{minipage}[b]{0.7\linewidth}
\subfigure[\small Error in the branch length estimation for distinct
depths of the branches.]{\label{fig:sixtaxa_br}
\includegraphics[scale=0.35]{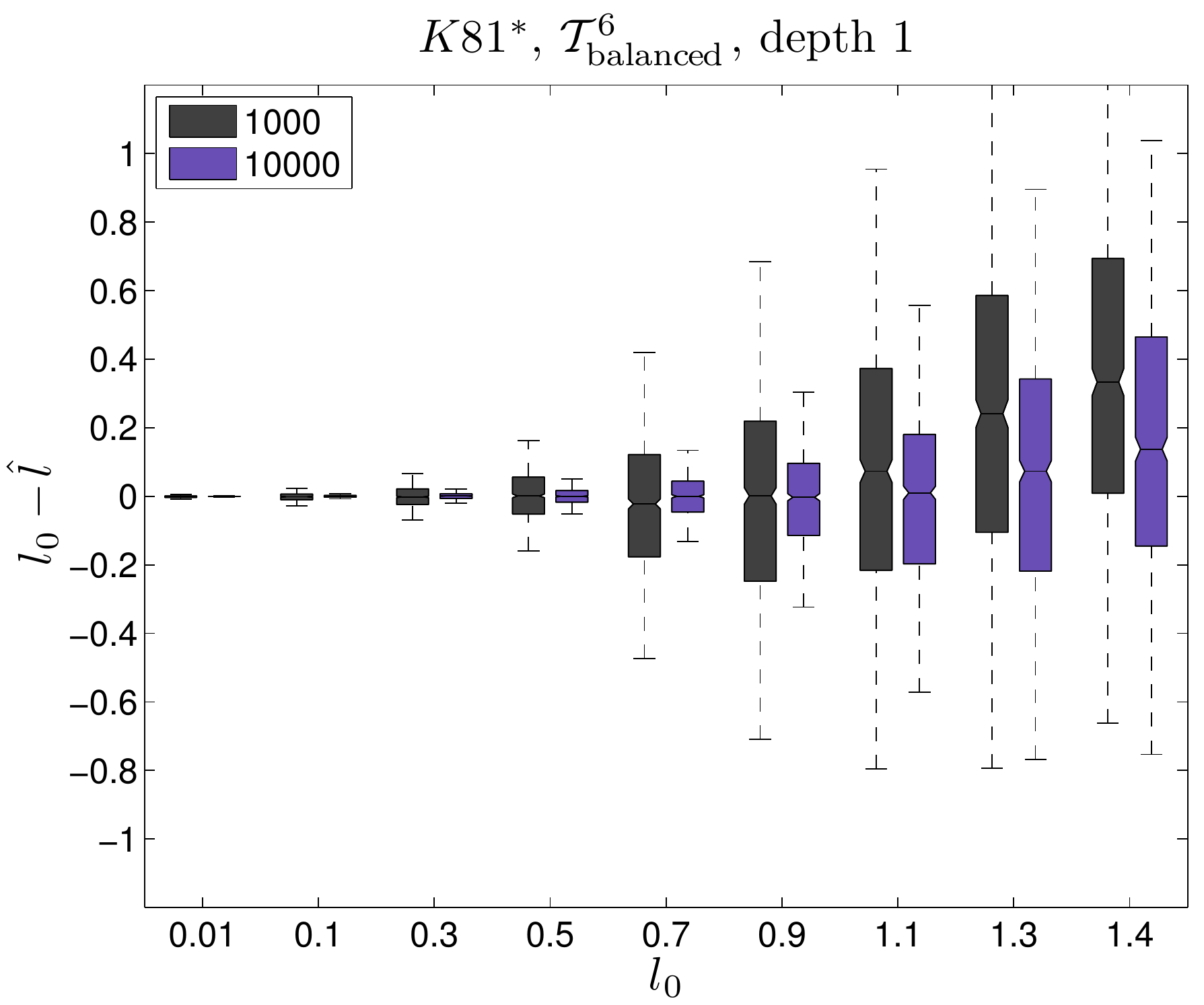}
\includegraphics[scale=0.35]{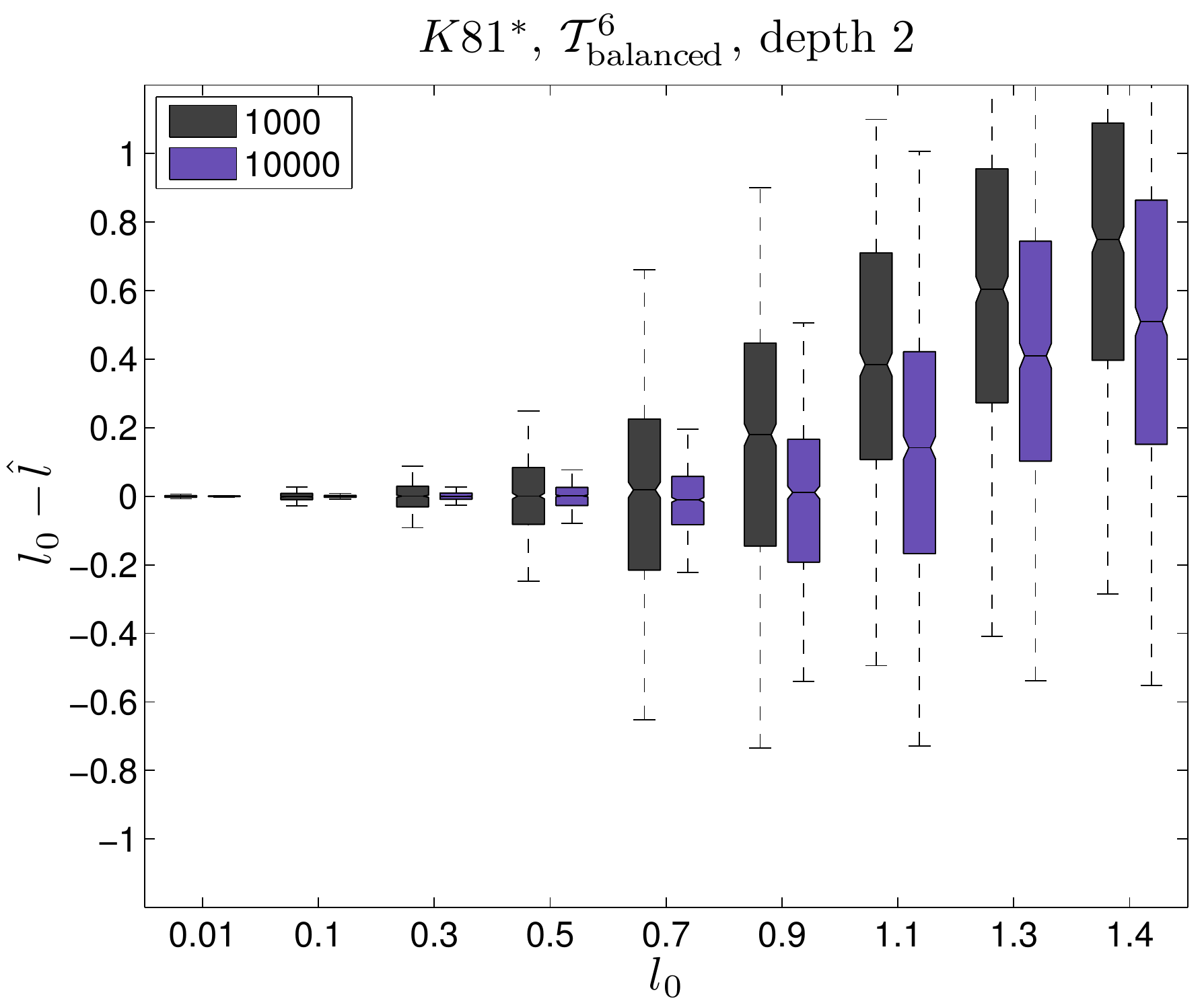}}
\subfigure[\small The average $\eltwo$ error between the original ($\xi$) and
estimated ($\hat{\xi}$) parameters.]{\label{fig:sixtaxa_dist}
\includegraphics[scale=0.35]{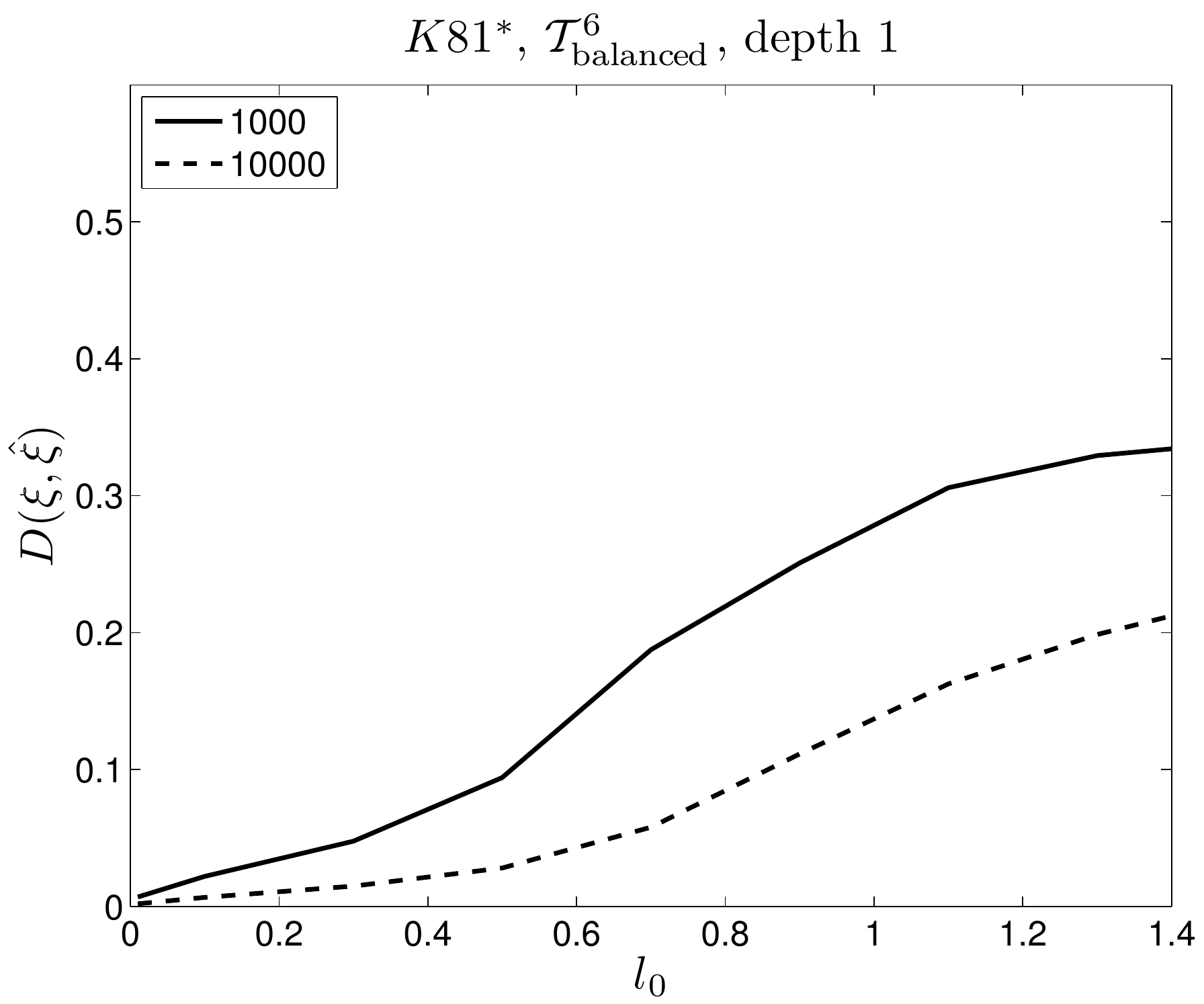}
\includegraphics[scale=0.35]{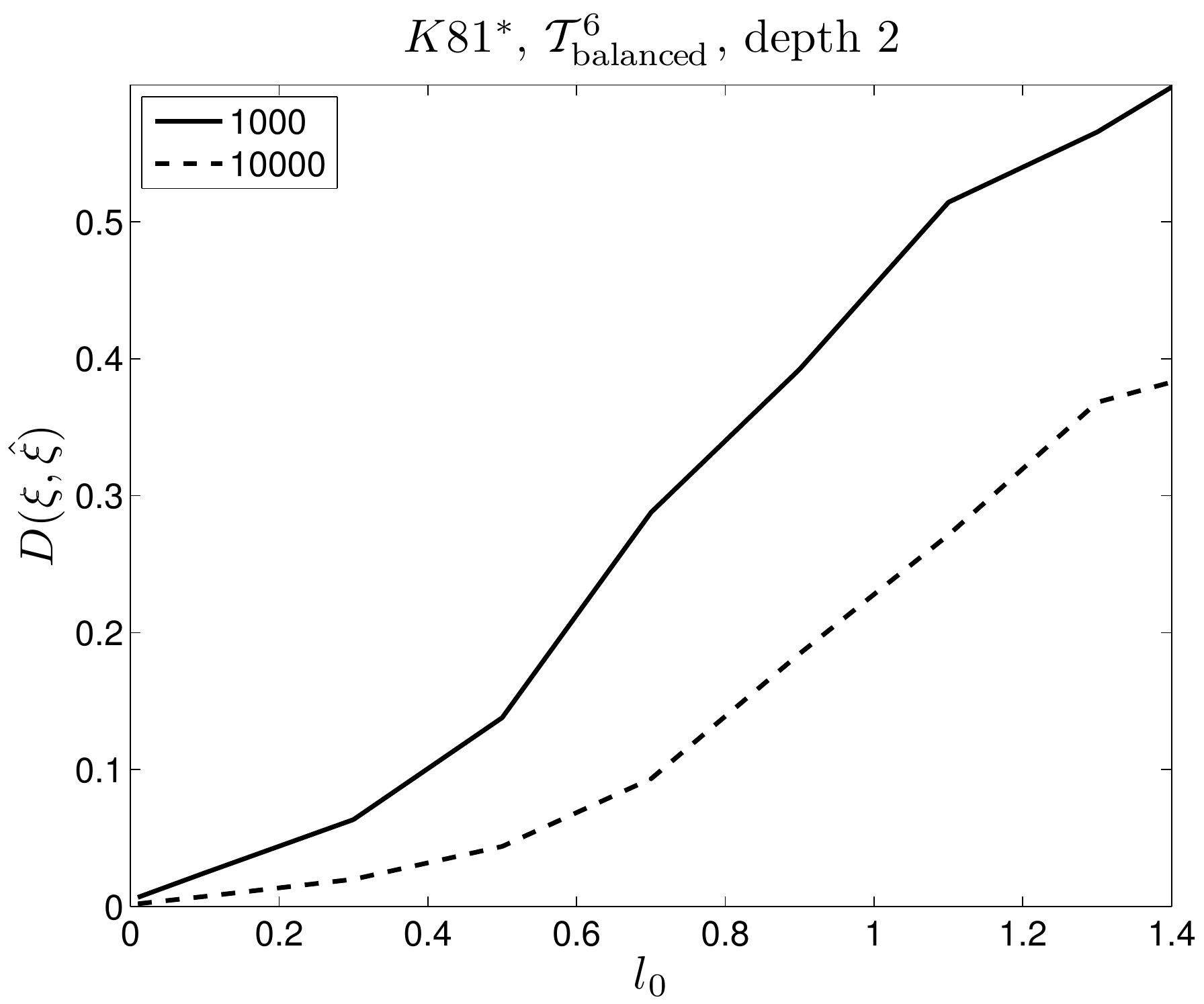}}
\subfigure[\small Distribution of the combined variance for distinct
depths of the branches. ]{\label{fig:sixtaxa_var}
\includegraphics[scale=0.35]{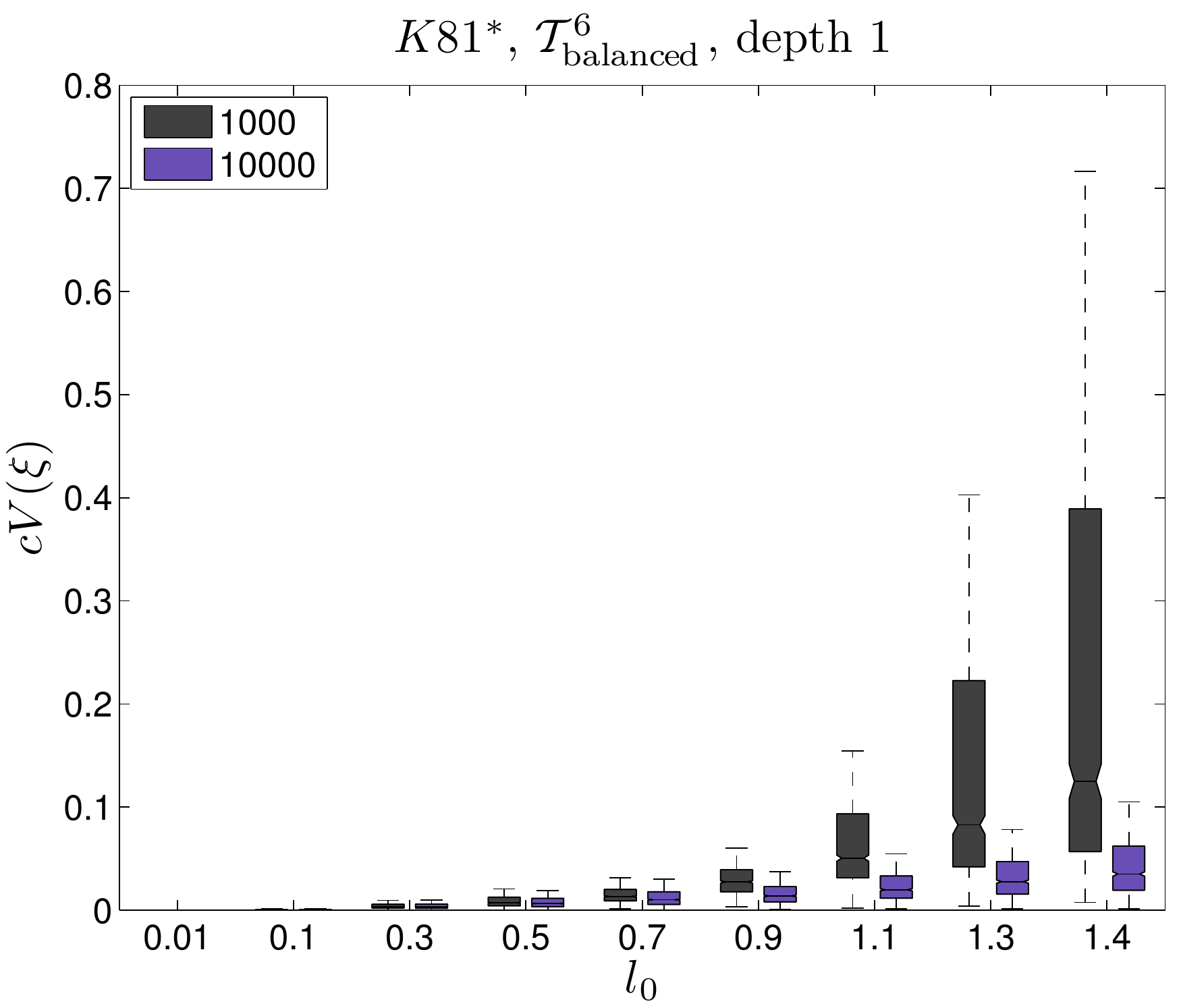}
\includegraphics[scale=0.35]{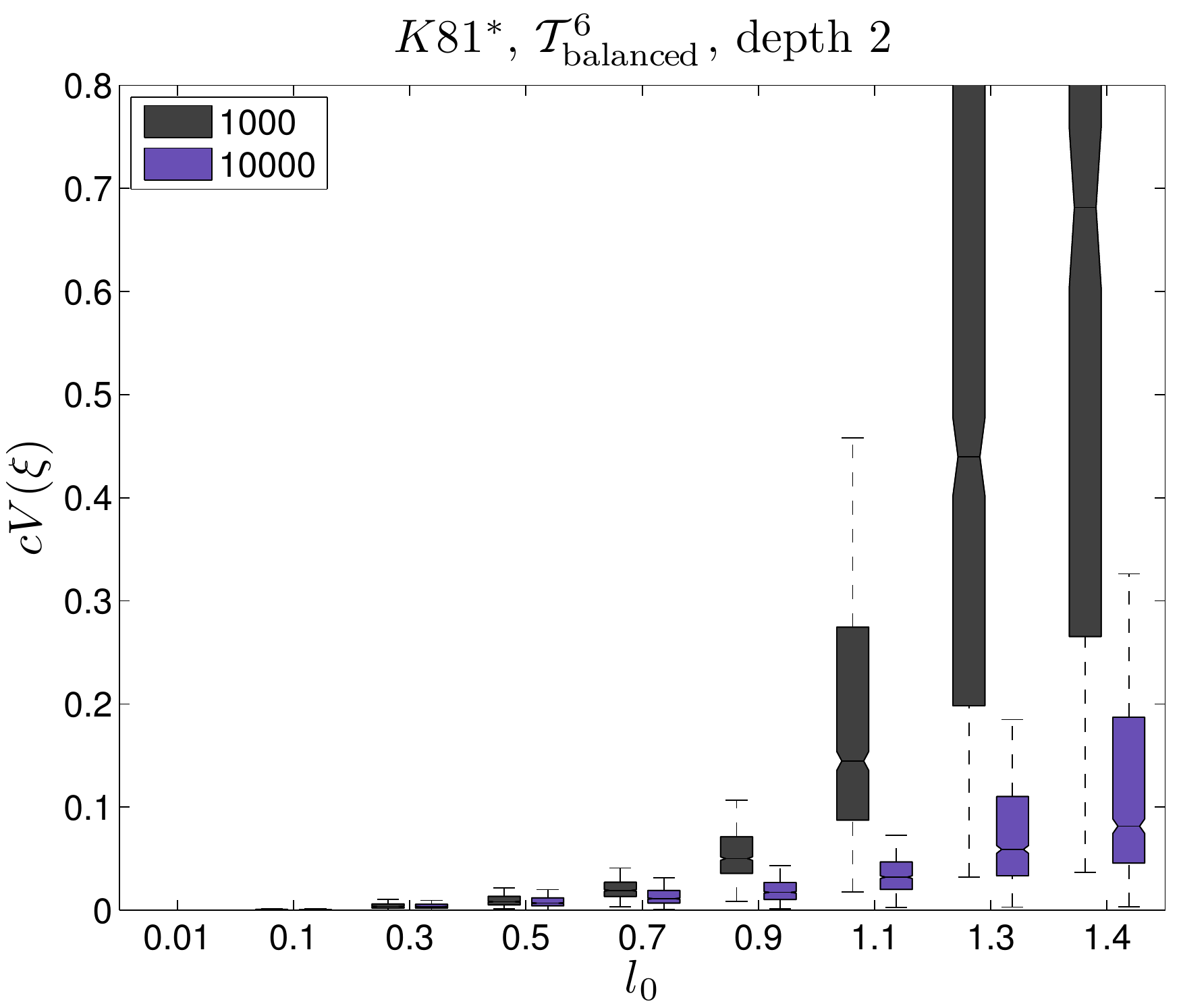}}
\end{minipage}
\caption{\small Results for the $1,000$ data sets generated on the
  $\sone^6$ tree for the $\kth$ model.}\label{fig:sixtaxa}
\end{figure}

\end{document}